\documentclass[twocolumn,superscriptaddress,aps,jcp,amsmath,amssymb]{revtex4-1}

\usepackage[latin9]{inputenc}
\usepackage{amsthm}
\usepackage{amsmath}
\usepackage{amssymb}
\usepackage{graphicx}
\usepackage{esint}
\usepackage{hyperref}
\usepackage{paralist}
\usepackage{xcolor} 
\usepackage{epstopdf}

\def\C{{\mathbb C}}% complex numbers
\def\R{{\mathbb R}}% real numbers
\def\RR{{\mathbb{R}}}
\def\le{\leqslant}% lessoreqal
\newcommand{\veps}{\varepsilon}

\newcommand{\wt}{\widetilde}
\newcommand{\ud}{\,\mathrm{d}}

\newcommand{\Or}{\mathcal{O}}

\theoremstyle{plain}

\theoremstyle{definition}

\newtheorem*{remark*}{Remark}

\newcommand{\mc}[1]{\mathcal{#1}}

\newcommand{\abs}[1]{\lvert#1\rvert}
\newcommand{\Abs}[1]{\left\lvert#1\right\rvert}

\newcommand{\ket}[1]{\lvert#1\rangle}

\newcommand{\FGA}{\text{FGA}}

\DeclareMathOperator{\tr}{tr}

\begin{document}

\title{Improved sampling and validation of frozen Gaussian
  approximation with surface hopping algorithm for nonadiabatic
  dynamics}

\author{Jianfeng Lu} \email{jianfeng@math.duke.edu}
\affiliation{Department of Mathematics, Duke University, Box 90320,
  Durham NC 27708, USA}\affiliation{Department of Physics and
  Department of Chemistry, Duke University, Durham NC 27708, USA}
\author{Zhennan Zhou} %\email{zhennan@math.duke.edu}
\affiliation{Department of Mathematics, Duke University, Box 90320,
  Durham NC 27708, USA}

\date{\today}

\begin{abstract}
  In the spirit of the fewest switches surface hopping, the frozen
  Gaussian approximation with surface hopping (FGA-SH) method samples
  a path integral representation of the non-adiabatic dynamics in the
  semiclassical regime. An improved sampling scheme is developed in
  this work for FGA-SH based on birth and death branching
  processes. The algorithm is validated for the standard test examples
  of non-adiabatic dynamics.
\end{abstract}

\maketitle

\section{Introduction}

The surface hopping algorithms, pioneered in \cite{TullyPreston:71}
and revamped as the fewest switches surface hopping (FSSH) algorithm
in \cite{Tully:90}, are widely used for mixed quantum-classical
dynamics in the non-adiabatic regime. The surface hopping algorithms
have been successfully applied to various scenarios where
non-adiabatic effect is important \cite{HammesSchifferTully:94,
  BarbaraMeyerRatner:96, Tully:98, Kapral:06, ShenviRoyTully:09,
  Barbatti:11, Subotnik:16}. Due to the huge popularity, the
development of surface hopping algorithms, which focuses on improving
the approximation to the exact quantum dynamics, further reducing the
computational cost, or taking into account the interaction with
environment, just to name a few, has been a very active research area
\cite{HammesSchifferTully:94, BarbaraMeyerRatner:96, PrezhdoRossky:97,
  Tully:98, HorenkoSalzmannSchmidtSchutte:02,
  JasperStechmannTruhlar:02, WuHerman:05,
  Bedard-HearnLarsenSchwartz:05, HannaKapral:05, Kapral:06,
  WuHerman:06, WuHerman:07, HannaKimKapral:07,
  SchmidtParandekarTully:08, ShenviRoyTully:09, Barbatti:11,
  SubotnikShenvi:11, LandrySubotnik:11, GorshkoviTretiakMozyrsky:13,
  SubotnikOuyangLandry:13, LandryFalkSubotnik:13, HannaKapral:13,
  JainHermanOuyangSubotnik:15, JainSubotnik:15, Subotnik:16, Kapral:16}. 

The underlying idea of the surface hopping algorithms is to use
classical trajectories with hopping between adiabatic energy surfaces
to approximate the exact Schr\"odinger dynamics, which is impractical
to solve directly due to the curse of dimensionality. While the
intuition behind the FSSH type algorithm is quite convincing, the
understanding of its systematic derivation from the exact
Schr\"odinger dynamics remains rather poor, despite huge progress in
recent years \cite{Herman:84, KapralCiccotti:99,
  HorenkoSalzmannSchmidtSchutte:02, HannaKapral:05, WuHerman:05,
  WuHerman:06, Kapral:06, GorshkoviTretiakMozyrsky:13,
  SubotnikOuyangLandry:13, Subotnik:16, Kapral:16}

In our previous work \cite{FGASH}, we gave a mathematically rigorous
derivation of a surface hopping type algorithm, called frozen Gaussian
approximation with surface hopping (FGA-SH), starting from the
nuclei-electron Schr\"odinger equation in the non-adiabatic regime.
The algorithm can be viewed as a natural extension of the Herman-Kluk
propagator \cite{HermanKluk, Kay:94, Kay:06}, which is a consistent
approximation to the single surface Schr\"odinger equation in the
semiclassical regime, to the non-adiabatic dynamics and hence takes
into account hopping between different energy surfaces.

The key observation behind the FGA-SH method is that the idea of
surface hopping leads to a path integral representation of the
semiclassical Schr\"odinger equations for multiple adiabatic surfaces
(referred to as the matrix Schr\"odinger equation in the sequel), for
which the path is given by the classical paths with hops between
adiabatic surfaces, in the same spirit of those used in FSSH. To
prevent any possible confusion, we emphasize that this is rather
different from the Feynman path integral, in particular, the paths are
similar to those trajectories of nuclei in FSSH and paths may carry
different weights. This path integral representation not only provides
a clear interpretation of the FSSH type algorithms in terms of a Monte
Carlo sampling scheme of the path integral, but also naturally leads
to further improvement of numerical schemes to approximate the path
integral, which the current work is devoted to.

The main new ingredient of the improved algorithm in this work is that
instead of using independent trajectories (as a direct Monte Carlo
sampling scheme), we reduce the sampling variance by adopting the
birth-death branching processes, which is typically used in other
context of path integral sampling, e.g., the diffusion Monte Carlo
algorithms \cite{GrimmStorer:69, GrimmStorer:71, Anderson:75}. As far
as we know, this is the first time such strategy is used for surface
hopping type algorithms, thanks to the novel path integral
interpretation. In fact, we expect the connection of the view point of
surface hopping algorithms with the path integral based methods (see
e.g., \cite{CaoVoth:94, SunMiller:97, JangVoth:99, Makri:99,
  Miller:01, CraigManolopoulos:04, LambertMakri:12,
  HabershonManolopoulosMarklandMiller:13}) would be quite fruitful and
would fertilize better algorithms for non-adiabatic
dynamics. The birth-death branching processes result in an
  adaptive pruning and splitting of surface hopping trajectories,
  which bears some similarity with the multiple spawning method
  \cite{MartinezBenNunLevine:96, BenNunMartinez:98} in the context of
  non-adiabatic dynamics, while the latter spawns Gaussian as a set of
  basis functions for a wave function approach, rather than
  semiclassical trajectories.

Besides the improved sampling scheme of FGA-SH, in this work we also
further elaborate the initial sampling of trajectories in the FGA-SH
method and also the calculation of observables based on averaging of
trajectories. We validate the improved FGA-SH method by various
numerical tests, which explore the dependence of the numerical error
on the semiclassical parameter $\veps$, the long time accuracy and
conservation of energy, the impact of the weighting factor, etc. for
the model test problems by Tully \cite{Tully:90} for nonadiabatic
dynamics.
  
The rest of the paper is organized as follows.  In Section
\ref{sec:method}, we review the path integral representation for
matrix Schr\"odinger equation in the semiclassical regime which leads
to the FGA-SH method introduced in \cite{FGASH}. The improved sampling
algorithm for FGA-SH is discussed in Section \ref{sec:algorithm}. We
validate the method by numerical tests in Section \ref{sec:numeric}. 
 The paper is concluded in
Section \ref{sec:conclusion}.

\section{Theory} \label{sec:method}
\subsection{Path integral representation for semiclassical matrix
  Schr\"odinger equations}

We consider the matrix Schr\"odinger equation with two electronic levels
in the adiabatic basis:
\begin{multline}\label{vSE}
  i \veps \partial_t  
  \begin{pmatrix} u_0 \\
    u_1 \end{pmatrix}  = -\frac{\veps^2}{2} \Delta_x \begin{pmatrix} u_0 \\
    u_1 \end{pmatrix} + 
  \begin{pmatrix} 
    E_0 \\
    & E_1
  \end{pmatrix} \begin{pmatrix} u_0 \\
    u_1 \end{pmatrix} \\ - \frac{\veps^2}{2} \begin{pmatrix}
    D_{00} & D_{01} \\
    D_{10} & D_{11}
  \end{pmatrix} \begin{pmatrix} u_0 \\
    u_1 \end{pmatrix} - \veps^2 \sum_{j=1}^m 
  \begin{pmatrix} 
    d_{00} & d_{01} \\
    d_{10} & d_{11}
  \end{pmatrix}_j
  \partial_{x_j} \begin{pmatrix} u_0 \\
    u_1 \end{pmatrix},
\end{multline}
where for $ k, l = 0, 1,\; j = 1, \ldots, m$,
\begin{align}
D_{kl}(x) =\langle \Psi_k(r;x), \Delta_x \Psi_l(r;x) \rangle_r, \\
\left(d_{kl}(x)\right)_j =\langle \Psi_k(r;x), \partial_{x_j}
\Psi_l(r;x) \rangle_r.
\end{align}
Here $m$ is the spatial dimension of the nuclei degree of freedom $x$
and $\Psi_k(r; x)$ are adiabatic states where $r$ denotes the
electronic degree of freedom. Note that our index convention of
$D_{kl}$ and $d_{kl}$ is flipped from that of Tully \cite{Tully:90}, we
choose the current convention so that the terms in \eqref{vSE} follows
the usual index convention of matrix-vector product.

The matrix Schr\"odinger equation is obtained from the nuclei-electron
Schr\"odinger equation by expanding the total wave function in the
adiabatic basis. After rescaling, the semiclassical nuclei-electron
Schr\"odinger equation reads (see e.g., \cite{Hagedorn:86, BO2})
\begin{equation}\label{eq:neSE}
i \veps \partial_t  \Phi(t,x,r)= - \frac{\veps^2}{2} \Delta_x \Phi (t,x,r)+H_e (r,x)\Phi(t,x,r),
\end{equation}
where $\Phi(t,x,r)$ denotes the nuclei-electron wave function and
$H_e(r,x)$ denotes electronic Hamiltonian (in a diabatic
representation).  The parameter $\veps$  is proportional to the square root of the ratio of
the electron mass to that of nuclei and is thus a small parameter (for
simplicity of notation, we have assumed that all nuclei share the same
mass). The adiabatic states $\Psi_k(r; x)$ are the eigenstates of
$H_e(r,x)$ with eigenvalues $E_k(x)$. Assume the first two adiabatic
states are well isolated from the rest of the states, we thus take the
following ansatz for the total wave function
\begin{equation}\label{eq:adiaexp}
\Phi(t,x,r)=u_0(t,x) \Psi_0(r;x)+u_1(t,x)\Psi_1(r;x).
\end{equation}
We obtain \eqref{vSE} by inserting \eqref{eq:adiaexp} into
\eqref{eq:neSE} and writing the equations in terms of $u_0$ and
$u_1$. While it is possible to generalize the method to take into
account of more than two adiabatic surfaces, we restrict to the case
of two for simplicity.

Solving the matrix Sch\"odinger equations \eqref{vSE} using conventional numerical methods is
impractical due to the potential high dimensionality of the nuclei
degree of freedom, and hence we resort to semiclassical methods which
exploit the limiting behavior of $\veps \to 0$. In previous work
\cite{FGASH}, we derived the frozen Gaussian approximation with
surface hopping (FGA-SH) from the matrix Schr\"odinger equation with
rigorous error bounds of the approximate nuclei wave function. The
algorithm follows the same spirit as the Tully's fewest switches
surface hopping (FSSH) method \cite{Tully:90}, except that the hopping
rule is different from FSSH, as will be explained in
subsection~\ref{sec:trajectory}.

The FGA-SH method can be understood as a path integral formulation of
the matrix Schr\"odinger equation in the spirit of surface hopping. It
approximates the solution $u = \bigl(\begin{smallmatrix} u_0 \\
  u_1 \end{smallmatrix}\bigr)$ as
\begin{equation}\label{eq:FGApathintegral}
  \begin{aligned}
  u(T, x) & = u_{\text{FGA-SH}}(T, x) + \Or(\veps) \\
  & = \mathbb{E}_{\wt{z}} \mc{F}\bigl(x; \{\wt{z}(s)\}_{0 \leq s \leq T}\bigr) + \Or(\veps), 
  \end{aligned}
\end{equation}
where the average is taken over an ensemble of trajectories we
describe below in Section~\ref{sec:trajectory} and the functional
$\mc{F}$ (expression given below in Section \ref{sec:functional})
depends on the trajectory $\wt{z}$ up to time $T$.  Our surface
hopping algorithm can be viewed as a Monte Carlo sampling scheme for
this path integral. As proved in \cite{FGASH}, $u_{\text{FGA-SH}}$
gives an approximation to the exact solution with error $\Or(\veps)$
(in $L^2$ metric) for any finite $T$.  For completeness, we provide in
the Appendix a brief discussion of the derivation of
\eqref{eq:FGApathintegral}. The readers may refer to \cite{FGASH} for
detailed asymptotic derivation.

\subsection{Surface hopping trajectories}\label{sec:trajectory}

Let us first specify the ensemble of trajectories used in
\eqref{eq:FGApathintegral}, which largely follows the spirit of
surface hopping trajectories.  The trajectory $\wt{z}(t)$ in
\eqref{eq:FGApathintegral} evolves on the extended phase space which
consists of the classical phase space on two energy surfaces: we write
$\wt{z}(t) = (z(t), l(t)) \in \R^{2m} \times \{0, 1\}$, where
$l(t) \in \{0, 1\}$ indicates the energy surface that the trajectory
lies on at time $t$.  The position and momentum $z(t) = (p(t), q(t))$
evolves by a Hamiltonian flow on the energy surface $l(t)$:
\begin{align}\label{eq:drift}
\begin{cases}
  & \dot{q}(t) = p(t); \\
  & \dot{p}(t) = - \nabla_q E_{l(t)}(q(t)). 
  \end{cases}
\end{align}
The trajectory hops between surfaces according to a Markov jump
process, with infinitesimal transition rate over the time period $(t, t + \delta t)$:
\begin{equation}\label{eq:infrate}
\mathbb{P}\bigl(l(t+ \delta t)=m \mid l(t)=n, \,z(t) = z \bigr) = \delta_{nm} + \lambda_{nm}(z) \delta t + o(\delta t) 
\end{equation}
for $m, n \in \{0, 1\}$, where the rate matrix is given by 
\begin{equation}
%  \lambda(z) = 
  \begin{pmatrix}
    \lambda_{00}(z) & \lambda_{01}(z) \\
    \lambda_{10}(z) & \lambda_{11}(z) 
  \end{pmatrix} =
  \begin{pmatrix}
    - \Abs{p \cdot d_{10}(q)} & \Abs{p \cdot d_{10}(q)} \\
    \Abs{p \cdot d_{01}(q)} & - \Abs{p \cdot d_{01}(q)}
  \end{pmatrix}.
\end{equation}
That is, if the trajectory is on the surface $0$ at time $t$, then
during the time interval $(t, t+\delta t)$, the probability that the
trajectory hops to the surface $1$ is given by
$\abs{p(t) \cdot d_{10}(q(t))} \delta t$ for sufficiently small
$\delta t$.  We remark that $p \cdot d_{10}(q)$ is in general complex,
and hence we take its modulus in the rate matrix; also note that the
rate is state dependent (on $z(t)$). The trajectory $\wt{z}(t)$ thus
follows a Markov switching process, which is piecewise deterministic.
The sampled trajectories follow the Hamiltonian dynamics on each
energy surface, with random hops to the other energy surfaces, and
thus are very similar in spirit to those trajectories used in the FSSH
method (though with different hopping rules).

% Equivalently, denote the probability distribution
% on the extended phase space at time $t$ by $f(t, z, l)$, the
% corresponding forward Kolmogorov equation is given by
% \begin{equation}
%   \frac{\partial}{\partial t} 
%   f(t, z, l) + 
%   \bigl\{ h_l, f(t, z, l) \bigr\} = \sum_{m = 0}^1  f(t, z, m) \lambda_{ml}(z),  
% \end{equation}
% where $\{\cdot, \cdot\}$ stands for the Poisson bracket corresponding to the Hamiltonian dynamics \eqref{eq:drift}, 
% \begin{equation*}
%   \bigl\{h, f\bigr\} = \partial_p h \cdot \partial_q f - \partial_q h \cdot \partial_p f,
% \end{equation*}

According to \eqref{eq:drift}, the position and momentum part
$z(t) = (p(t), q(t)$) of the trajectory $\wt{z}(t) $ is
continuous and piecewise differentiable, while $l(t)$ is piecewise
constant with almost surely finite many jumps during any finite time
interval. Given a realization of the trajectory
$\wt{z}(t) = (z(t), l(t))$ starting from initial condition
$\wt{z}(0) = (z(0), l(0))$, we denote by $n$ the number of jumps $l(t)$
has in the time interval $[0, T]$ (thus $n$ is a random variable) and
also the discontinuity set of $l(t)$ as
$\bigl\{t_1, \cdots, t_n\bigr\}$, which is an increasingly ordered
random sequence. At each of those time, the trajectory switches from
one energy surface to the other; and thus $t_k$, $k = 1, \ldots, n$,
are referred to as \emph{hopping times} in the sequel.

The starting point of the trajectory $\wt{z}(0) = (z(0), l(0))$ is
sampled according to a distribution on the extended phase space
determined by the initial condition of the matrix Schr\"odinger
equation. Given the initial wave function $u_k(0,x)$ for $k = 0,1$, we
denote the Gaussian packet transform of $u$ as
\begin{equation}\label{eq:initA}
  A_0(z, l)  = 2^{\frac m 2}\int_{\R^m} u_{l}(0, y) e^{\frac{i}{\veps} (-p_0\cdot(y-q)+ \frac{i}{2}|y-q|^2)} \ud y. 
\end{equation}
Then the $\wt{z}(0) = (z(0), l(0))$ is sampled from the measure
$\mathbb{P}_0(z(0),l(0))$ with probability density on $\R^{2m} \times
\{0, 1\}$ proportional to $\abs{A_0(z(0), l(0))}$. Here we assume that
$A_0(\cdot, k)$ is integrable on $\RR^{2m}$ for each $k = 0, 1$ and we
denote the normalizing factor as
\begin{equation}\label{eq:z}
  \mathcal{Z}_0 = \frac{1}{(2\pi\veps)^{3m/2}}\sum_{k=0,1} \int_{\R^{2m}} \Abs{A_0(z, k)} \ud z. 
\end{equation}
Thus, the initial point of the trajectory $\wt{z}(t)$ follows
  the distribution
\begin{equation}
  \mathbb{P}_0(z, l) = \mathcal{Z}_0^{-1} \frac{1}{(2\pi \veps)^{3m/2}} \abs{A_0(z, l)}.
\end{equation}
We will discuss the numerical sampling of initial points in
Section~\ref{sec:init_sample}. 

%%%%%%%%%%%%%%%%%%%%%%

\subsection{Ensemble average of surface hopping trajectories}\label{sec:functional}

Given the description of the ensemble of paths $\wt{z}(t)$, $t \in [0, T]$, we now specify the
functional $\mc{F}$ in the path integral
\eqref{eq:FGApathintegral}. Recall that we denote $n$ the number of
hops of the trajectory and $\bigl\{t_1, \cdots, t_n\bigr\}$ the
sequence of hopping times. For convenience, we also denote $t_0 = 0$
and $t_{n+1} = T$, the starting and final time of the trajectory.
% We also use the short hand
% $T_{k:1}=(t_k,\cdots,t_1)$ to denote the first $k$ hopping times $k =
% 0, \ldots, n$, ordered backwardly, with the convention that $T_{0:1}$
% denotes an empty set. 
The functional $\mc{F}$ is then given by
%\begin{widetext}
\begin{multline}\label{eq:defF}
  \mc{F}\bigl(x; \{\wt{z}(t)\}_{0 \leq t \leq T}\bigr) = \ket{l(T)} \,  \frac{\mathcal{Z}_0}{\Abs{A_0(\wt{z}(0))}}
  A(T) \times \\
  \times \exp\Bigl(\frac{i}{\veps} \Theta(T, x)\Bigr)
  \exp\Bigl(w(T)\Bigr) \prod_{k=1}^n
  \frac{\tau^{(k)}}{\Abs{\tau^{(k)}}}\,,
\end{multline}
where $\ket{0} = \bigl( \begin{smallmatrix} 1 \\ 0 \end{smallmatrix}
\bigr)$ and $\ket{1} = \bigl( \begin{smallmatrix} 0 \\
  1 \end{smallmatrix} \bigr)$
denotes the electronic state associated with each surface, $\mc{Z}_0$
is defined in \eqref{eq:z}, $A_0(\wt{z}(0))=A_0(z(0),l(0))$ is given
in \eqref{eq:initA}, and all the other terms depend on the trajectory,
in particular, the sequence of hopping times (we suppress the
dependence in the notation for simplicity). An outline of the
  argument leading to \eqref{eq:defF} is provided in the Appendix,
  with the detailed asymptotic derivation provided in \cite{FGASH}. It
  comes from a stochastic representation of a high dimensional
  integral involving all possible hopping times of a surface hopping
  trajectory.

Let us explain the terms appeared in
  \eqref{eq:defF}. $\ket{l(T)}$ gives the electronic state of the
  trajectory at time $T$, and the factor
  $\mc{Z}_0 / \abs{A_0(\wt{z}(0))}$ results from the initial
  sampling. The term
\begin{equation*}
  A(T) \exp\Bigl(\frac{i}{\veps} \Theta(T, x)\Bigr)
\end{equation*}
resembles the familiar amplitude $(A(T))$  and phase $(\Theta(T, x))$ expression from the Herman-Kluk
propagator \cite{HermanKluk, Kay:94, Kay:06, SwartRousse}, in particular, the phase term $\Theta$ takes
the following form
\begin{equation}
  \Theta(t, x) = S(t)
  + \frac{i}{2} \abs{x-q(t)}^2 + p(t) \cdot ( x-q(t) ),
\end{equation}
where $S(t)$ is the classical action associated with the
  trajectory and recall that $z(t) = (p(t), q(t))$ is the momentum and
  position of the trajectory.  The amplitude $A$ and action $S$ solve
the ODEs
\begin{align}
  \dot{S}(t) & =\frac{1}{2} p(t)^2 - E_{l(t)} (q(t)), \label{eq:S}\\
  \dot{A}(t) & = \frac{1}{2}
              A \tr\Bigl( Z(t)^{-1}\bigl(\partial_z p(t)
              - i \partial_z q(t) \nabla^2_q E_{l(t)}(q(t)) \bigr) \Bigr)
              \nonumber \\
             & \quad - A\, d_{l(t)l(t)}(q(t)) \cdot p(t).
\end{align} 
with initial conditions $S(0;\wt{z}(0))=0$ and
$A(0;\wt{z}(0))=A_0(\wt{z}(0))$.  Here $\partial_z$ is short for
$\partial_z = \partial_{q(0)} - i \partial_{p(0)}$ and
$Z(t) = \partial_z(q(t) + i p(t))$. Those equations are similar
  to the evolution equations in Herman-Kluk propagator, except that
  similar to the evolution equations for $(p(t), q(t))$, the above
  ODEs also depend on the current surface $l(t)$ of the trajectory.
Also note that it is clear for the ODEs that the value $S(t)$ and
$A(t)$ are determined by the trajectory up to time $t$.  

The last term in
\eqref{eq:defF} involves the hopping coefficients $\tau^{(k)}$ at each hopping, 
given by
\begin{equation}\label{eq:deftau}
  \tau^{(k)} = - p(t_k) \cdot d_{l(t_k^+) l(t_k^-)}(q(t_k)),
\end{equation}
where $l(t_k^-)$ and $l(t_k^+)$ give the energy surface index before
and after the hop at hopping time $t_k$ (so that
$l(t_k^-)\ne l(t_k^+)$). Recall that this is exactly related to
  the jumping intensity used for surface hopping of the trajectories.
  Since $\tau^{(k)}$ is complex valued in general, we have chosen its
modulus as the jumping intensity $\lambda$ in the surface hopping
trajectory, the term $\tau^{(k)} / \Abs{\tau^{(k)}}$ in
\eqref{eq:defF} is used to correct the phase factor due to the complex
value. 

Finally, the weighting factor $w$ in \eqref{eq:defF} solves the ODE
\begin{equation}\label{eq:defw}
  \dot{w}(t) = 
  \Abs{ p(t) \cdot d_{(1 - l(t))l(t)}(q(t))}, 
\end{equation}
with initial condition $w(0) = 0$. Thus, it is the accumulated jumping
intensity of the trajectory. The appearance of the weighting term in
\eqref{eq:defF} is due to the fact that the infinitesimal transition
rate \eqref{eq:infrate} of the trajectory $\wt{z}$ is non-homogeneous
and depends on the current position and momentum. Therefore, we need
to reweight the terms in the path-integral representation such that
the average gives the correct wave function. Without the weighting
term, the path integral formulation is no longer an accurate
approximation \cite{FGASH}; we also show in Section
\ref{sec:weighting} the crucial role of such weighting terms for
calculating observables associated to the non-adiabatic dynamics.

For the algorithmic purpose, we remark that we can combine $A$ with the weighting factor $w$ as 
\begin{equation}\label{eq:defgamma}
  \Gamma(t) = \frac{A(t)}{|A(0)|} \exp \bigl( w(t) \bigr), 
\end{equation}
which solves the ODE 
\begin{multline}\label{eq:Gamma}
  \dot{\Gamma}(t) = \frac{1}{2} \Gamma \tr\Bigl(
  Z(t)^{-1}\bigl(\partial_z p(t) - i \partial_z q(t) \nabla^2_q
  E_{l(t)}(q(t)) \bigr) \Bigr)
  \\
  + \Gamma \Bigl(\Abs{ p(t) \cdot d_{(1 - l(t))l(t)}(q(t))} - 
  p(t) \cdot d_{l(t)l(t)}(q(t))  \Bigr)
\end{multline}
with initial condition $\Gamma(0;\wt{z}(0)) =
A_0(\wt{z}(0))/|A_0(\wt{z}(0))|$. The quantity $\Abs{\Gamma(t)}$ will be treated as the weight of the trajectory in our algorithm. Thus we will prune trajectories with small weight, and branch trajectories with larger weights to reduce the variance of the stochastic sampling algorithm.

\section{Algorithm}\label{sec:algorithm}

\subsection{FGA-SH sampling based on birth/death branching process} 

As discussed before, the path integral representation readily suggests
a direct Monte Carlo sampling strategy: An ensemble of independent
trajectories are generated as in section~\ref{sec:trajectory} and then
the average of $\mc{F}$ is calculated as in
section~\ref{sec:functional} to approximate the time-dependent wave
function or the associated observables. This is the algorithm used in
our previous work \cite{FGASH}. Here we present an improved sampling
strategy based on birth/death branching process to reduce the sampling
variance of the ensemble average, borrowing a familiar variance
reduction method in the context of diffusion Monte Carlo
algorithms. The basic idea is to prune the trajectories with small
weights, while duplicate those with larger weights
($\abs{\Gamma(t)}$), in a consistent way that the mean is preserved,
while avoiding few trajectories carry huge weights so to reduce the
variance of the sampling. Note that the trajectories are no longer
independent, the dependence comes in due to the branching step.

The algorithm starts by sampling a collection of initial points for
the trajectories and estimate $\mc{Z}_0$ defined in \eqref{eq:z},
which will be discussed in more details in
  section~\ref{sec:init_sample}. We denote by $M(0)$ the number of
trajectories initially. Each trajectory carries the information of
position $q_{\alpha}$, momentum $p_{\alpha}$, the index of energy
surface $l_{\alpha}$, classical action $S_{\alpha}$, and a weight
$\gamma_{\alpha}$; initially we set $S_{\alpha} = 0$ and
$\gamma_{\alpha} = A_0(\wt{z}_{\alpha}(0))/|A(\wt{z}_{\alpha}(0))|$
for each trajectory. We use $\gamma_{\alpha}$ here to distinguish with
$\Gamma$ since during the branching process, the weight
$\gamma_{\alpha}$ of a trajectory will be redistributed among the
offspring and hence differ from $\Gamma$, as will be discussed in the algorithm below. 

The propagation of the trajectories are carried out as follows: For
each time step of size $\Delta t$, the following steps are performed
in order:
\setdefaultleftmargin{1.5em}{.5em}{}{}{}{}
\begin{enumerate}
\item Evolve the position and momentum $p(t), q(t)$ by the Hamiltonian
  dynamics \eqref{eq:drift} on the current surface $l(t)$ for time
  interval $\Delta t$ for each trajectory (we omit the index of
  trajectory in the algorithm description for simplicity of notation).

\item Evolve the quantities $S$ and $\gamma$ following \eqref{eq:S}
  and \eqref{eq:Gamma} respectively according to the current surface
  of the trajectory $l(t)$. Note that as discussed in
  section~\ref{sec:functional}, the quantities $S$ and $\gamma$ at
  time $t$ only depend on the portion of the trajectory $\wt{z}$ up to
  time $t$. Therefore we may calculate them for each trajectory on the
  fly.  These ODEs can be numerically solved by standard ODE
  integrators, for example, a $4$-th order Runge-Kutta scheme is
  chosen in our implementation.

\item Hopping attempts. The probability that a surface hop occurs within
  the time interval $(t, t + \Delta t)$ is given by
  $\Delta t \abs{\lambda_{(1 - l(t))l(t)}}$. For $\Delta t$
  sufficiently small, we can neglect the event that two hops happen
  within the time interval. With this probability, the trajectory is
  hopped to the other surface, so that the label of the energy surface
  is changed: $l(t + \Delta t) = 1 - l(t)$ and the phase change
  $ \tau / \abs{\tau}$ is recorded.

\item Birth/death branching. For every $N_{\text{branch}}$ steps we do
  the branching according to the weight that the trajectories carry at
  time $t + \Delta t$: $\gamma = \gamma(t + \Delta t)$. 
  For each trajectory, we generate a random number
  $\xi$ uniformly distributed on $[0, 1]$. Denoting $[\abs{\gamma}]$
  as the integer part of $\abs{\gamma}$, the birth/death is given by
  \begin{compactitem}
  \item If $\xi> \abs{\gamma} - [\abs{\gamma}]$, we replace the
    current trajectory with $[\abs{\gamma}] + 1$ trajectories
    identical to the parent trajectory except that the weight is reset
    to be $\gamma/|\gamma|$; 
  \item If $\xi\leq \abs{\gamma} - [\abs{\gamma}]$, we replace the
    current trajectory with $[\abs{\gamma}]$ trajectories identical to
    the parent trajectory except that the weight is reset to be
    $\gamma/|\gamma|$. If $[\abs{\gamma}] = 0$, this means that we kill
    the parent trajectory without any offsprings.
  \end{compactitem}
  Note that the branching rule is designed so that on average the
  total weight of the offsprings is equal to the weight of the parent
  trajectory, whereas the weight of each offspring is of order $1$.
\end{enumerate}

After propagating the trajectories till $t=T$, the  wave function can be reconstructed following \eqref{eq:FGApathintegral} with the modification to take into account the birth/death branching process. The path integral is approximated by 
\begin{multline}\label{eq:defFbd}
  u_{\text{FGA-SH}}(T, x) = \frac{\mathcal{Z}_0}{M(0)} \sum_{\alpha = 1}^{M(T)}
  \ket{l_{\alpha}(T)} \,
  \gamma_{\alpha}(T) \times \\
  \times \exp\Bigl(\frac{i}{\veps} \Theta_{\alpha}(T, x)\Bigr)
  \prod_{k=1}^{n_{\alpha}} \frac{\tau_{\alpha}^{(k)}}{\Abs{\tau_{\alpha}^{(k)}}}, 
\end{multline}
where $M(T)$ denotes the number of trajectory at time $T$ and we use
subscript $\alpha$ explicitly to emphasize the dependence of these
quantities on the right hand side on each trajectory.  We also remark
that $|\gamma_{\alpha}| \approx 1$ due to the branching process, so it
mainly contributes to a phase factor in the summation.

We emphasize that, except for the step of birth/death branching, these
is no exchange of information between the trajectories, and moreover,
the branching history does not contribute to the modified average of
trajectories \eqref{eq:defFbd}, and is hence not necessary to be
stored.  Therefore, the computational cost to generate one trajectory
in the current algorithm is almost the same as those with fully
independent trajectories, e.g., the direct Monte Carlo sampling of the
path integral \cite{FGASH}.

\subsection{Initial sampling} \label{sec:init_sample}

We now further elaborate the initial sampling of the trajectories.
For simplicity, let us assume that initially, the wave function is
only non-zero on surface $0$, so that the trajectories will all
initiate on that surface. Recall that we aim to sample $z(0)$
according to the probability distribution
\begin{equation}\label{eq:target}
\mathbb P_0(q,p, 0) = \frac{1}{ \mathcal{Z}_0}  \frac{1}{(2\pi\veps)^{3m/2}} \left|A_0(z(0), 0)\right|,
\end{equation}
where $A_0$ is defined in \eqref{eq:initA} and $\mc{Z}_0$ in
\eqref{eq:z}.  We assume that $A_0(z(0), 0)$ can be obtained with some
accuracy. Explicit calculation is unfortunately only possible in low
dimensions for general initial data, due to the curse of dimension in
numerical quadrature. A special case is if we assume $u_0(y)$ is a
Gaussian wave packet, for which $A_0$ and $\mc{Z}_0$ can be obtained
explicitly. We now give the expression of $\mathbb{P}_0$ for Gaussian
initial data, as this is used in our numerical tests. For example, if
we consider a semiclassical wave packet centered at $y=\tilde q$ with
momentum $\tilde p$,
\[
u_0(y)= \exp \left(\frac{i}{\veps} \tilde p \cdot (y-\tilde q )\right)
\exp\left(-\sum_{j=1}^m \frac{a_j}{2\veps} (y_j-\tilde q_j)^2 \right),
\]
where $a_j$ are positive constants. By direct calculation, we have 
\begin{widetext}
\begin{equation}\label{eq:Cex}
  A_0(q,p, 0) = 2^{m}(\pi \veps)^{\frac m 2} \prod_{j=1}^m (1+a_j)^{-\frac 1 2} \exp \left(- \frac{(\tilde p_j-p_j)^2+a_j(\tilde q_j-q_j)^2}{2(1+a_j)\veps} \right)\exp \left(\frac{i (a_j\tilde q_j+q_j)(\tilde p_j-p_j)}{(1+a_j)\veps}+ \frac{i \left(p_j q_j -\tilde p_j \tilde q_j\right)}{\veps} \right),
\end{equation}
\end{widetext}
and thus 
\begin{multline}\label{eq:Cexab}
  \Abs{A_0(q,p, 0)} = 2^{m}(\pi \veps)^{\frac m 2} \prod_{j=1}^m (1+a_j)^{-\frac 1 2}  \\
  \times \exp \left(- \frac{(\tilde p_j-p_j)^2+a_j(\tilde q_j-q_j)^2}{2(1+a_j)\veps} \right).
\end{multline}
The resulting probability  distribution for the initial position and momentum is a multivariate normal distribution centered at $q=\tilde q$ and $p=\tilde p$ with standard derivation $\mathcal O(\sqrt \veps)$. The normalizing constant $\mc{Z}_0$ can be also calculated analytically as
\begin{equation*}
  \mathcal{Z}_0 = \frac{1}{(2\pi\veps)^{3m/2}} \int_{\R^{2m}} \Abs{A_0(z, 0)} \ud z 
  =2^{\frac m 2} \prod_{j=1}^m\left( \frac{1+a_j}{a_j} \right).
\end{equation*}
Therefore, the probability distribution \eqref{eq:target} is obtained
explicitly.  In the special case that $a_j=1$, $\forall\, j$, we have
$\mathcal Z_0=2^m$ and the initial probabilistic measure is given by
\[
\mathbb P (p,q, 0)=2^{-2m}(\veps\pi)^{-m} \exp \left(- \frac{|\tilde
    p-p|^2+|\tilde q-q|^2}{4\veps} \right).
\]

The initial momentum and position of the trajectories
$(p_{\alpha}(0), q_{\alpha}(0))$ are sampled independently from the
distribution and we set $l_{\alpha} = 0$, $S_{\alpha} = 0$ and the
initial weight
\[
\gamma_\alpha = \frac{A_0 (z_{\alpha}(0), 0)}{|A_0(z_{\alpha}(0), 0)|}.
\]
For the Gaussian initial data, we can directly sample $(p, q)$ from
$\mathbb{P}_0$, which is a Gaussian distribution on the phase
space. For more general $\mathbb{P}_0$, some importance sampling might
be needed, which we will not go into details here.

\subsection{Computing observables}

For many applications, the goal is not to approximate the wave
function itself, but rather the time evolution of certain
observable. For the purpose of calculating an observable, it is
  often the case that we do not need to reconstruct the wave function
  on a mesh, which is not feasible in high dimensions anyway.

We take the mass on each energy surface as an example, denoted by
\begin{equation}
  m_k= \int_{\R^m} \left| {u_k}(T, x) \right|^2   \ud x, \quad k=0,1,
\end{equation}
other macroscopic quantities can be treated in a similar fashion.
From the approximation of the wave function in FGA-SH, we have
%\begin{widetext}
\begin{equation*}
\left| {u_k}(T, x) \right|^2 = {u_k}(T, x) \bar{u}_k(T, x). 
\end{equation*}
Using \eqref{eq:defFbd} for both $u_k$ and the complex conjugate $\bar{u}_k$, we arrive at
\begin{align}
\left| {u_k}(T, x) \right|^2
&\approx \frac{ \mathcal{Z}_0^2}{(M(0))^2}   \sum_{\alpha, \beta=1}^{M(T)} \Biggl[ \delta_{k,l_\alpha(T)} \delta_{k,l_\beta(T)} \gamma_\alpha \overline{\gamma_\beta}  \times \nonumber \\
& \quad \times  g_{\alpha,\beta}(x) \prod_{k=1}^{n_{\alpha}} \frac{\tau_{\alpha}^{(k)}}{\Abs{\tau_{\alpha}^{(k)}}}   \prod_{k=1}^{n_{\beta}} \frac{ \bar{\tau}_{\beta}^{(k)}}{\Abs{\tau_{\beta}^{(k)}}}  \Biggr], \label{eq:usquare}
\end{align}
%\end{widetext}
where $\alpha$ and $\beta$ are indices for two trajectories (for $u_k$ and $\bar{u}_k$ respectively), and we
have use the short-hand notation 
\begin{align*}
g_{\alpha,\beta}(x) & = \exp\Bigl(\frac{i}{\veps} \Theta_\alpha-\frac{i}{\veps} \bar{\Theta}_\beta\Bigr).
\end{align*}
We remark that while the formula \eqref{eq:usquare} appears to complex, the resulting $m_k$ is in fact real due to the symmetry of interchanging indices $\alpha$ and $\beta$. 

We observe that the only dependence on $x$ of the right hand side of
\eqref{eq:usquare} is through $g_{\alpha,\beta}(x)$, and since it is a
Gaussian function, the integration in $x$ can be carried out
analytically:
\begin{multline*}
  \int_{\R^m} g_{\alpha,\beta}(x) \ud x = (\pi \veps)^{\frac m 2}
  e^{\frac{i}{\veps} (S_{\alpha}(t) - S_{\beta}(t))} \times \\
  \times e^{- \frac{i}{2\veps} (q_\alpha(t) - q_{\beta}(t))\cdot
    (p_\alpha(t)+p_\beta(t))} \times \\
  \times e^{-\frac{1}{4\veps}\abs{q_\alpha(t)
      -q_\beta(t)}^2-\frac{1}{4\veps}\abs{p_\alpha(t) - p_\beta(t)}^2}, 
\end{multline*}
Therefore, given the sampled FGA-SH trajectories (and hence
$p_{\alpha}(t), q_{\alpha}(t)$ etc.), we can estimate $m_k$ without
reconstruction and numerical quadrature of the wave functions.

% \begin{widetext}
% \begin{equation}
% m_k= \int_{\R^m} \left| {u_k}(T, x) \right|^2   \ud x \approx  \frac{ \mathcal{Z}_0^2}{(M(0))^2}   \sum_{\alpha=1}^{M(T)}  \sum_{\beta=1}^{M(T)}\Biggl[ \delta_{k,n_\alpha} \delta_{k,n_\beta}\gamma_\alpha \overline{\gamma_\beta}   \prod_{k=1}^{n_{\alpha}} \frac{\tau_{\alpha}^{(k)}}{\Abs{\tau_{\alpha}^{(k)}}}   \prod_{k=1}^{n_{\beta}} \frac{ \overline{\tau_{\beta}^{(k)}}}{\Abs{\tau_{\beta}^{(k)}}} \int_{\R^m} g_{\alpha,\beta}(x) \ud x  \Biggr].
% \end{equation}
% \end{widetext}

\section{Results and discussions} \label{sec:numeric}
\subsection{Tully's three examples} \label{sec:Tully}

We now revisit the standard models examples similar to
  those of Tully's \cite{Tully:90}, which are test cases widely used
for non-adiabatic dynamics. 
The algorithm is not limited to one
  dimension, the comparison for higher dimensional systems (for
  instance the spin-boson system) will be considered in future works.
All the examples below have two adiabatic states, which means that the
Hilbert space corresponding to the electronic degree of freedom is
equivalent to $\C^2$, and hence the electronic Hamiltonian $H_e$ (in a
diabatic representation) is equivalent to a $2\times 2$ matrix.

\noindent\textbf{Example 1. Simple avoided crossing.}  

For the simple avoided crossing example, we will vary the energy gap
(controlled by the small parameter $\delta$) between the two surfaces
as $\veps \to 0$, which is a more interesting scenario. For this, we
consider a class of electronic Hamiltonian given by a product of a
scalar function $F^\delta(x)$ and a $2 \times 2$ matrix $M(x)$
independent of $\delta$, namely
\begin{equation*}
H_e^{\delta}(x)= F^\delta(x) M(x).
\end{equation*}
Since $F^{\delta}$ is a scalar, $H_e^{\delta}$ and $M$ share the same
eigenfunctions. Denoting the eigenvalues of $M$ by $\lambda_k$, we
have
\begin{equation}\label{case1}
E_k^{\delta}(x) =F^\delta (x)  \lambda_k(x).
\end{equation}
Then, we obtain that, for $k \ne l$,
\[
d^\delta_{lk}=\frac{\langle \Psi_l, \nabla_x H_e^\delta \Psi_k \rangle}{E_k^{\delta}-E_l^{\delta}} = \frac{ \langle \Psi_l, \nabla_x M \Psi_k \rangle}{\lambda_k-\lambda_l}, 
\]
and similarly, 
\[
D^\delta_{lk}=\langle \Psi_l, \Delta_x \Psi_k \rangle=\frac{\langle \Psi_l, \Delta_x M \Psi_k \rangle-2 \nabla_x \lambda_k \cdot d_{kl}}{\lambda_k-\lambda_l}.
\]
Therefore, $d^\delta_{lk}$ and $D^\delta_{lk}$ are independent of
$F^{\delta}$ (and of $\delta$ in particular), so we thereby suppress
the superscript $\delta$. While we can take $F^{\delta}$ such that
energy surfaces become close and even touch each other as
$\delta \to 0$.

\begin{figure}
\includegraphics[scale=0.275]{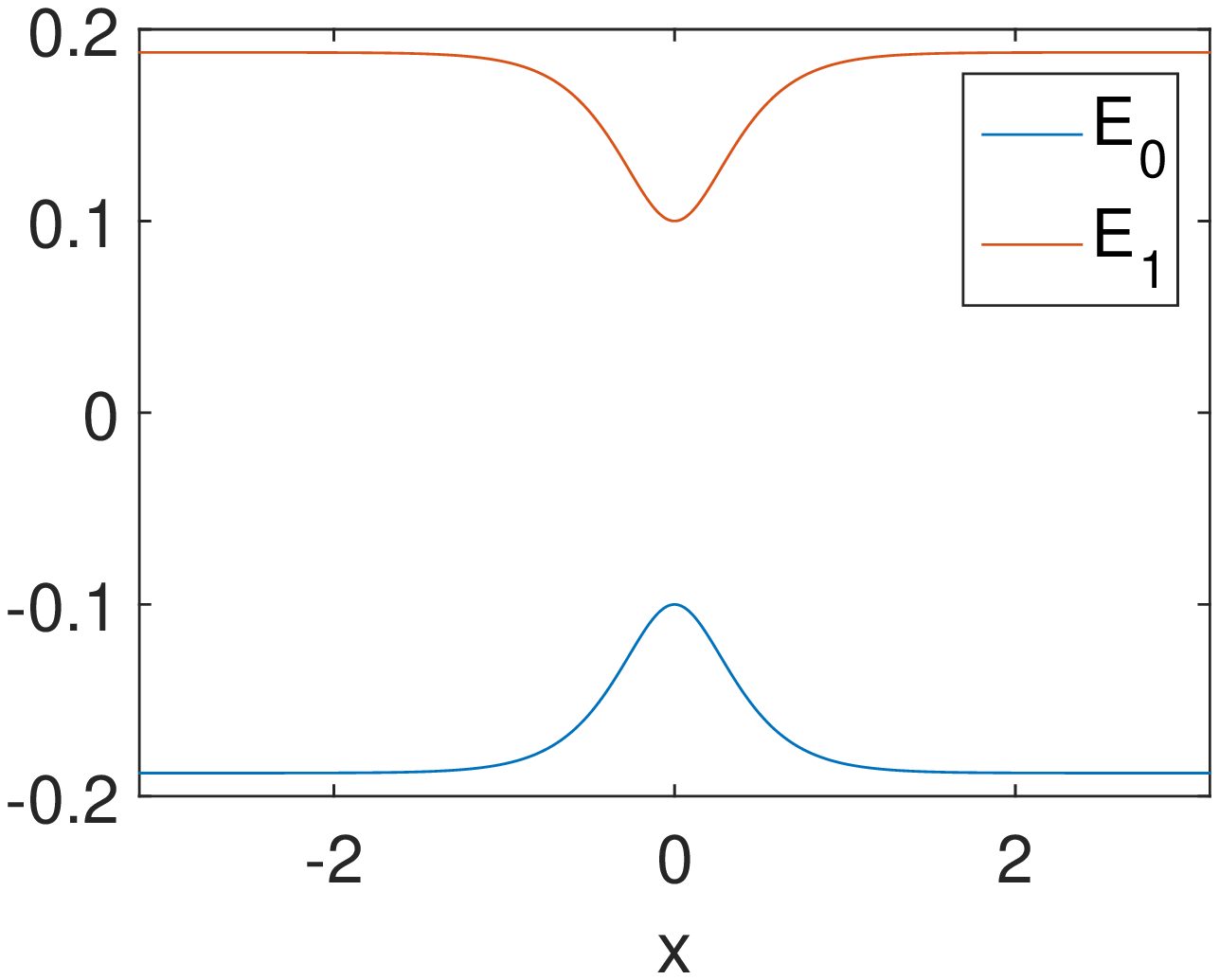}\includegraphics[scale=0.275]{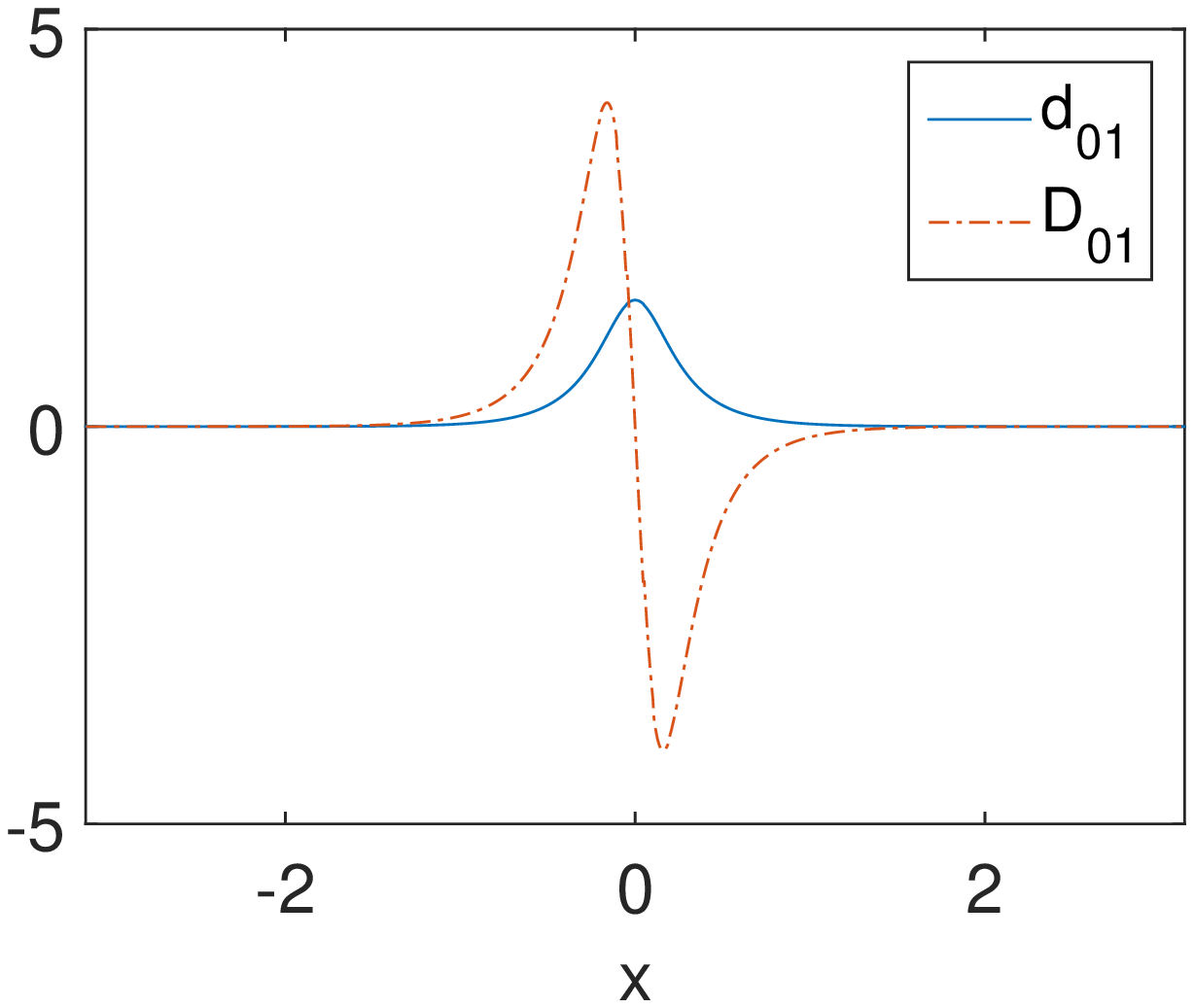} \\
\caption{Example 1. Left: Eigenvalues of $M$ (related to the
  eigenvalues of $H_e^{\delta}$ by $F^{\delta}$). Right: the coupling
  vectors of $H_e^{\delta}$, which are invariant with respect to
  $\delta$ (see text for further discussion). }
\label{fig:a2}
\end{figure}

For the simple avoided crossing, we choose $M$ to be
\[
M=
\begin{pmatrix}
\frac{\tanh (wx)}{2\pi}& 0.1\\
 0.1 & -\frac{\tanh (wx)}{2\pi}
\end{pmatrix},
\]
where $w$ is a stretching parameter. The eigenvalues of $M$ are \[
\pm \sqrt{\frac{\tanh^2 (w x)}{4\pi^2}+0.01 }, 
\] 
plotted in Figure \ref{fig:a2}.  We observe that the two eigenvalue
surfaces are close around $x=0$. By the plots of $d_{10}$ and $D_{10}$
in Figure \ref{fig:a2}, we see the coupling is significant around
$x=0$ as well. To control the energy gap, we introduce the following
$F^\delta$ function,
\[
F^\delta (x)=C_g ( 1 +(\delta -1 ) e^{- x^2}),  
\] 
such that $F^\delta(x) = \mathcal{O}(\delta)$ around $x=0$ and
$F^{\delta}(0) = \delta$, so that the energy gap vanishes at $x = 0$
as $\delta \to 0$. Here $C_g$ is a parameter we will use to adjust the
global energy gap of the adiabatic surfaces. We introduce
  $F^\delta$ to get a family of examples for different $\veps$, which
  facilitates the study of the algorithm for different semiclassical
  parameter $\veps$, as in Section~\ref{sec:diffveps}. We will focus on the most interesting regime of parameter choice
  $\delta=\mathcal O(\veps)$, so that the energy gap is comparable to the semiclassical parameter. 

\begin{figure}
\includegraphics[scale=0.275]{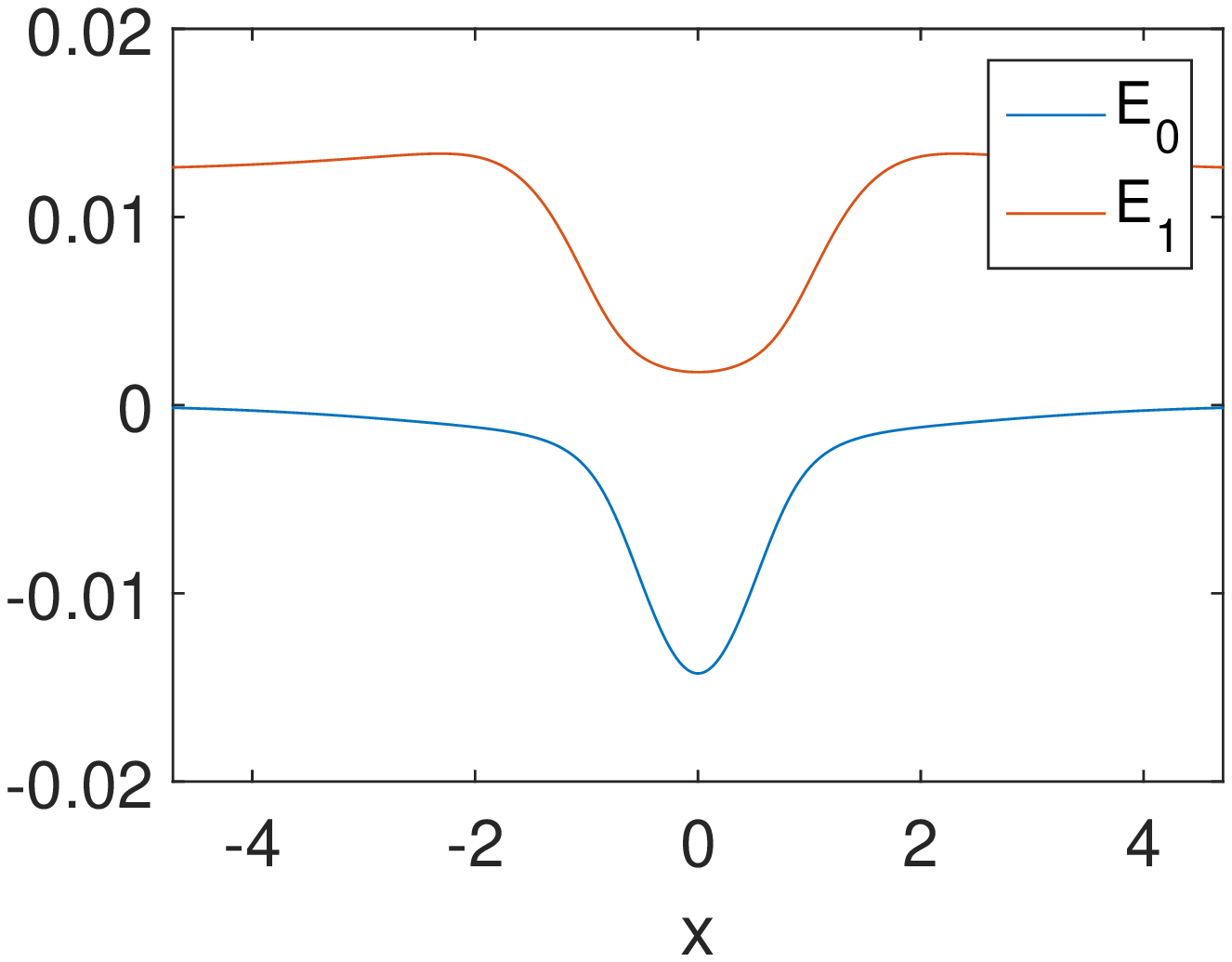}\includegraphics[scale=0.275]{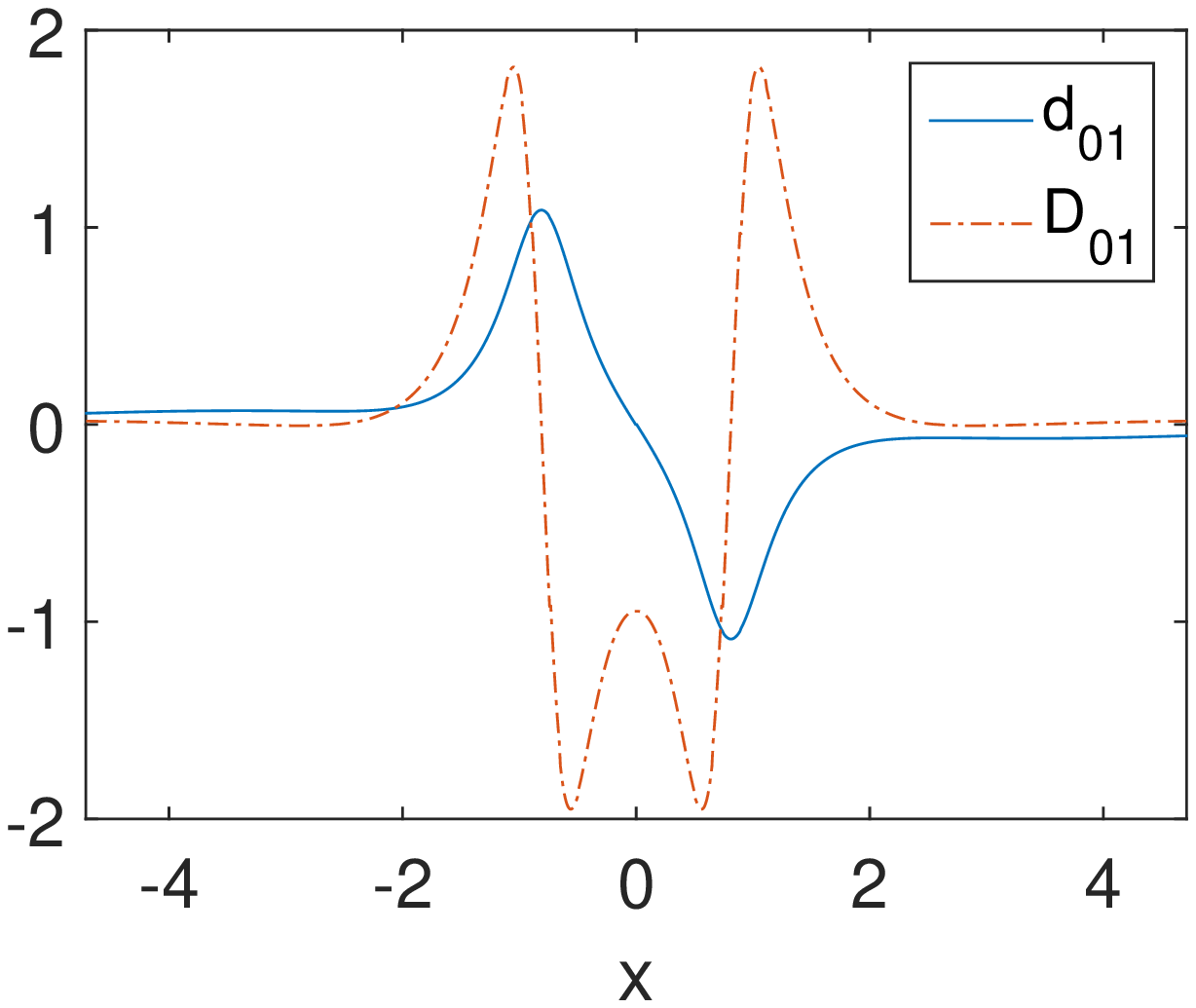} \\
\caption{Example 2. Left: Eigenvalues of $H_e$. Right:
  the coupling vectors of $H_e$.}
\label{fig:b2}
\end{figure}

\noindent\textbf{Example 2. Dual avoided crossing.} We choose $H_e$ to be
\[
H_e=\frac{1}{20}
\begin{pmatrix}
0 & 0.1 e^{-0.06{x^2}} \\
 0.1 e^{-0.06 x^2} & -0.5e^{-{x^2}}+0.25
\end{pmatrix}. 
\]
Observe that, the two eigenvalues are closest to each other when  $x=\pm 1$. The coupling vectors  around these points $x=\pm 1$ are significantly larger than their values elsewhere as shown in Figure \ref{fig:b2}. This explains why the model is often referred to as the dual avoided crossing.
The energy surfaces and coupling vector are illustrated in Figure \ref{fig:b2}.

\begin{figure}
\includegraphics[scale=0.275]{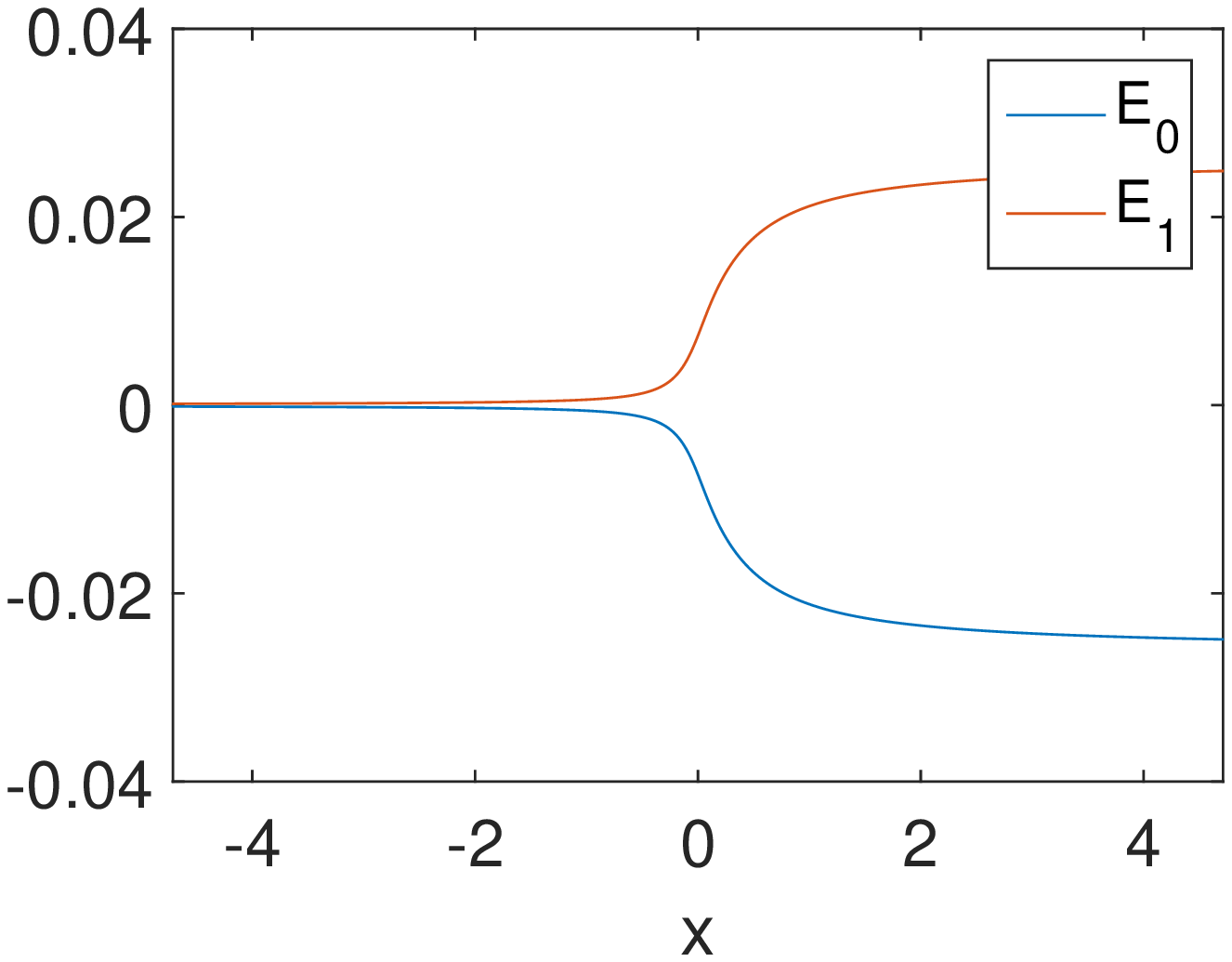}\includegraphics[scale=0.275]{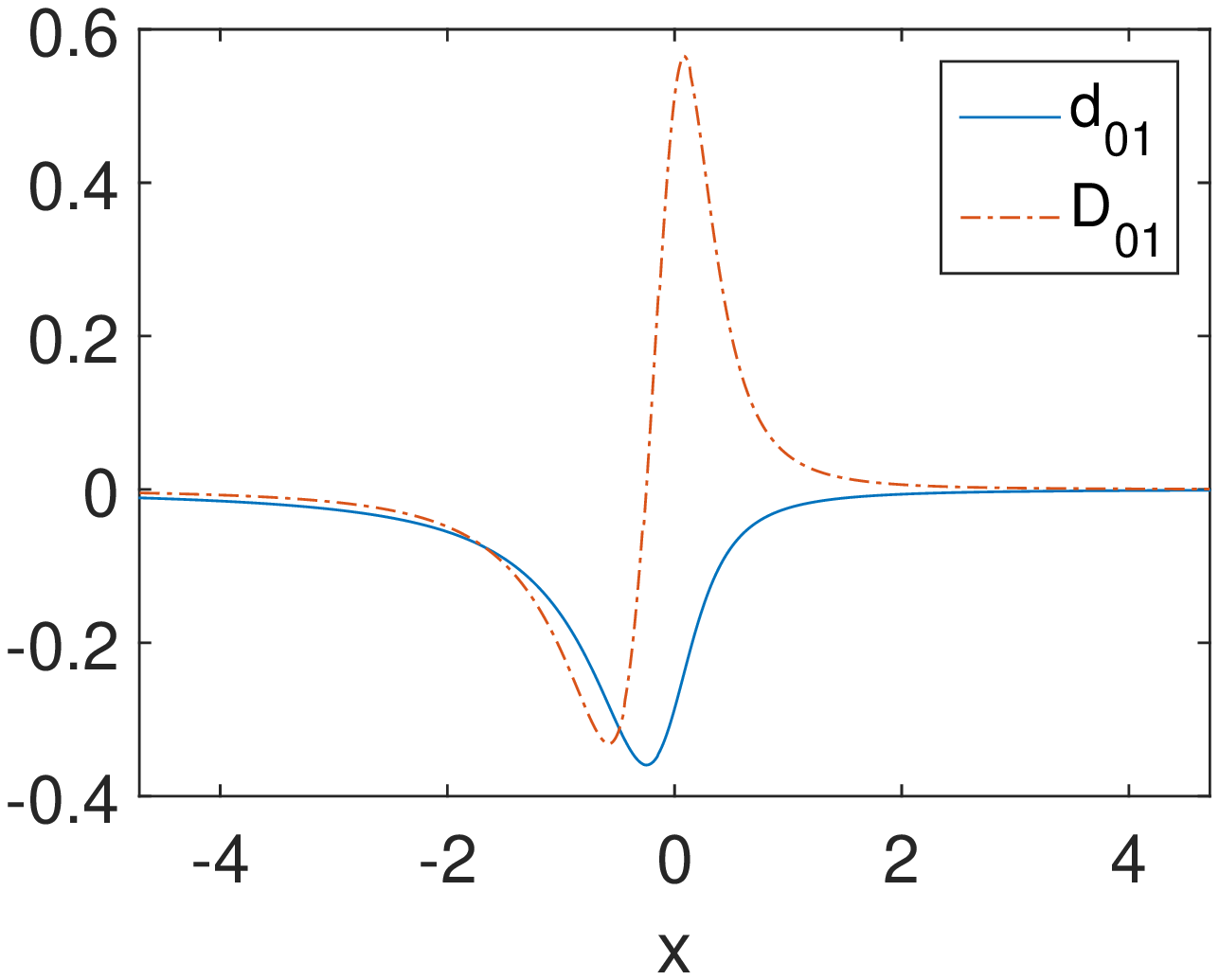} \\
\caption{Example 3. Left: Eigenvalues of $H_e$. Right:
  the coupling vectors of $H_e$.}
\label{fig:c2}
\end{figure}

\noindent\textbf{Example 3. Extended coupling with reflection.} 
In this example, $H_e$ is set to be 
\[
H_e=F
\begin{pmatrix}
\frac{1}{20}  &\frac{1}{20} \left(\arctan(2x) +\frac{\pi}{2} \right) \\
\frac{1}{20} \left(\arctan(2x) +\frac{\pi}{2} \right) & -\frac{1}{20} 
\end{pmatrix},
\]
where
\[
F=\frac{1}{20} \left( \arctan(5x)+\frac{\pi}2+\delta \right).
\]
Hence, as $x\rightarrow \infty$, the eigenvalues of $H_e$,
$E_{0/1}(x) \rightarrow \mp \pi^2/400$, and as
$x\rightarrow -\infty$, the eigenvalues of $H_e$, $E_{0/1}(x)
\rightarrow \mp 0$. As shown in Figure \ref{fig:c2}, this model
involves an extended region of strong nonadiabatic coupling when
$x<0$. Moreover, as $x>0$, the upper energy surface is repulsive so
that it is likely that trajectories moving from left to right on the
excited energy surface will be reflected while those on the ground
energy surface will be transmitted.
The energy surfaces and coupling vectors are illustrated in Figure \ref{fig:c2}.
Compared to  Tully's extended coupling example in \cite{Tully:90}, we slightly modify the potential energy surface so that the eigenvalues $E_{0/1}(x)$ as functions of $x$ become smooth.

\subsection{Convergence test with various $\veps$}\label{sec:diffveps}

\begin{figure}
\includegraphics[scale=0.55]{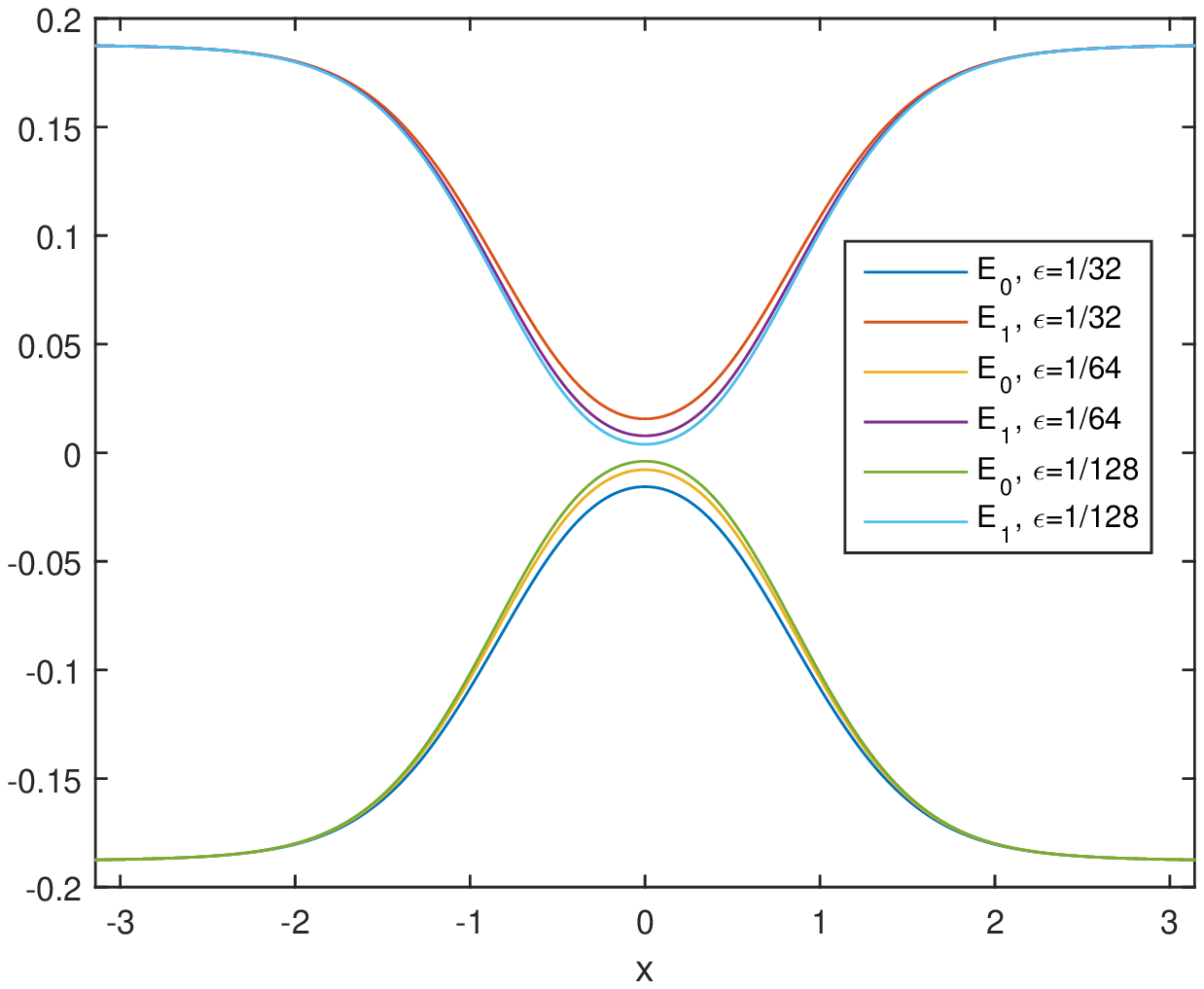} \\
\includegraphics[scale=0.55]{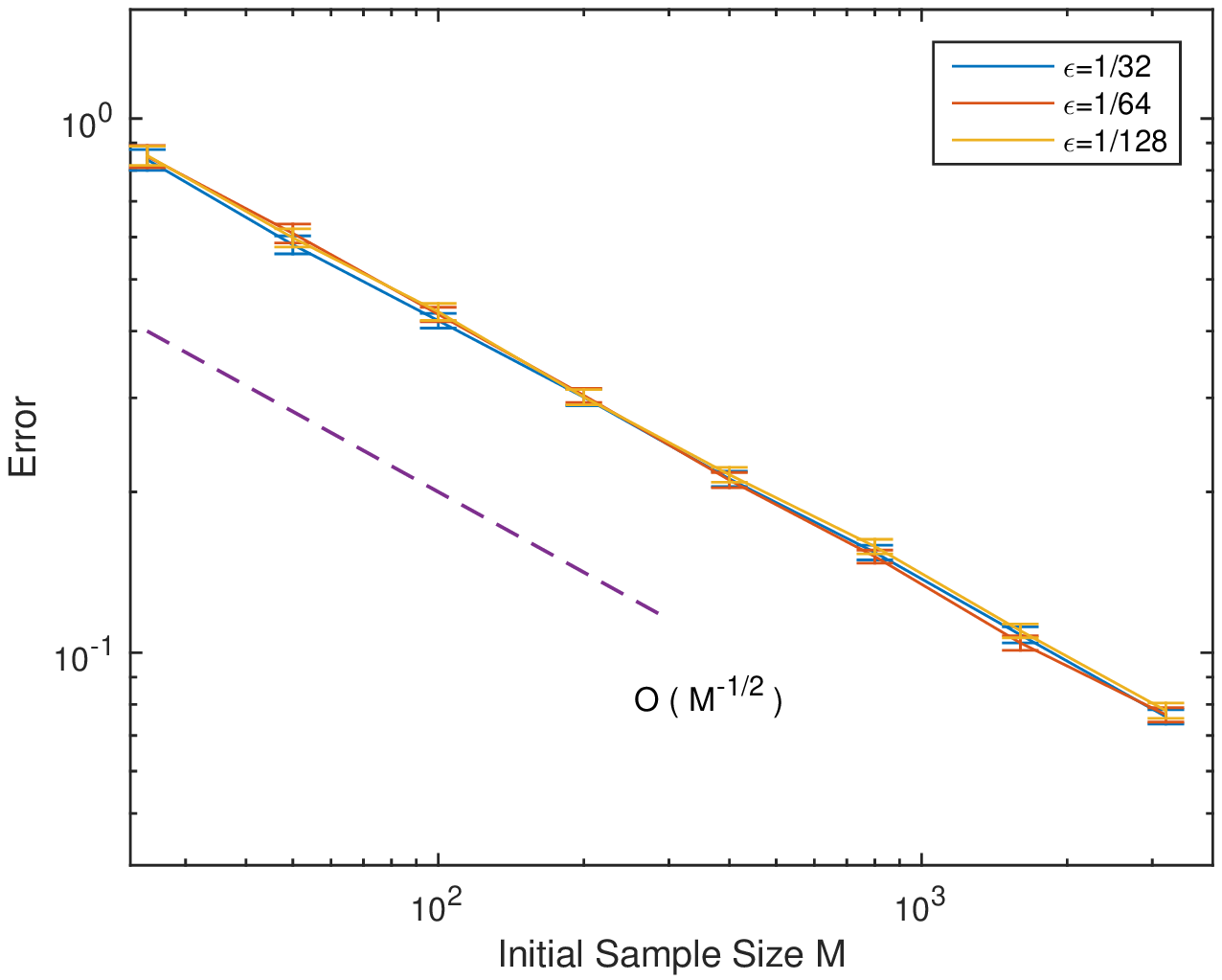} \\
\caption{Top: Eigenvalues of $H_e$ for $\veps=\frac{1}{32}$,
  $\frac{1}{64}$ and $\frac{1}{128}$. Bottom: Empirical average of the
  error in the wave functions with confidence intervals.}
\label{fig:test1}
\end{figure}

In this test, we test the algorithm on Example 1 for
$\veps=\frac{1}{32}$, $\frac{1}{64}$ and $\frac{1}{128}$ with $w=2$,
$\delta=5 \veps$ and $C_g=1$. The purpose is to understand the
performance of the algorithm for systems with different semiclassical
parameter $\veps$.  The corresponding energy surfaces are plotted in
Figure \ref{fig:test1}, while the coupling vectors stay unchanged. The
initial condition is chosen concentrated on the lower surface only,
and assumed to be the Gaussian wave packet
\begin{equation}\label{eq:u00}
u_0(0,x)= (32\veps)^{-1/4} e^{\frac{i}{\veps}k_0 \cdot (x-y_0)} e^{- \frac{1}{2\veps}(x-y_0)^2},
\end{equation}
where $y_0$ and $k_0$ denote the initial position and momentum of the
semiclassical wave packet respectively, the packet is normalzied so
that the $L^2$ norm is $1$.  In the set of tests, we fix $y_0=-1.5$
and $k_0=1.5$. For the initial trajectory number $M=25$, $50$, $100$,
$200$, $400$, $800$, $1600$ and $3200$, the numerical test is repeated
for $100$ times in each case, and empirical average of the $L^2$
errors of the wave functions with confidence intervals are plotted in
Figure \ref{fig:test1}. We observe that, the numerical error is
dominated by the stochastic sampling error which is of order
$\mathcal O(M^{-1/2})$, which is consistent with the Monte Carlo
nature of the algorithm. The stochastic sampling error also appears to
be rather independent of the small parameter $\veps$.

\subsection{Error growth in time and conservation of energy in a
  bouncing back test}

\begin{figure}
\includegraphics[scale=0.55]{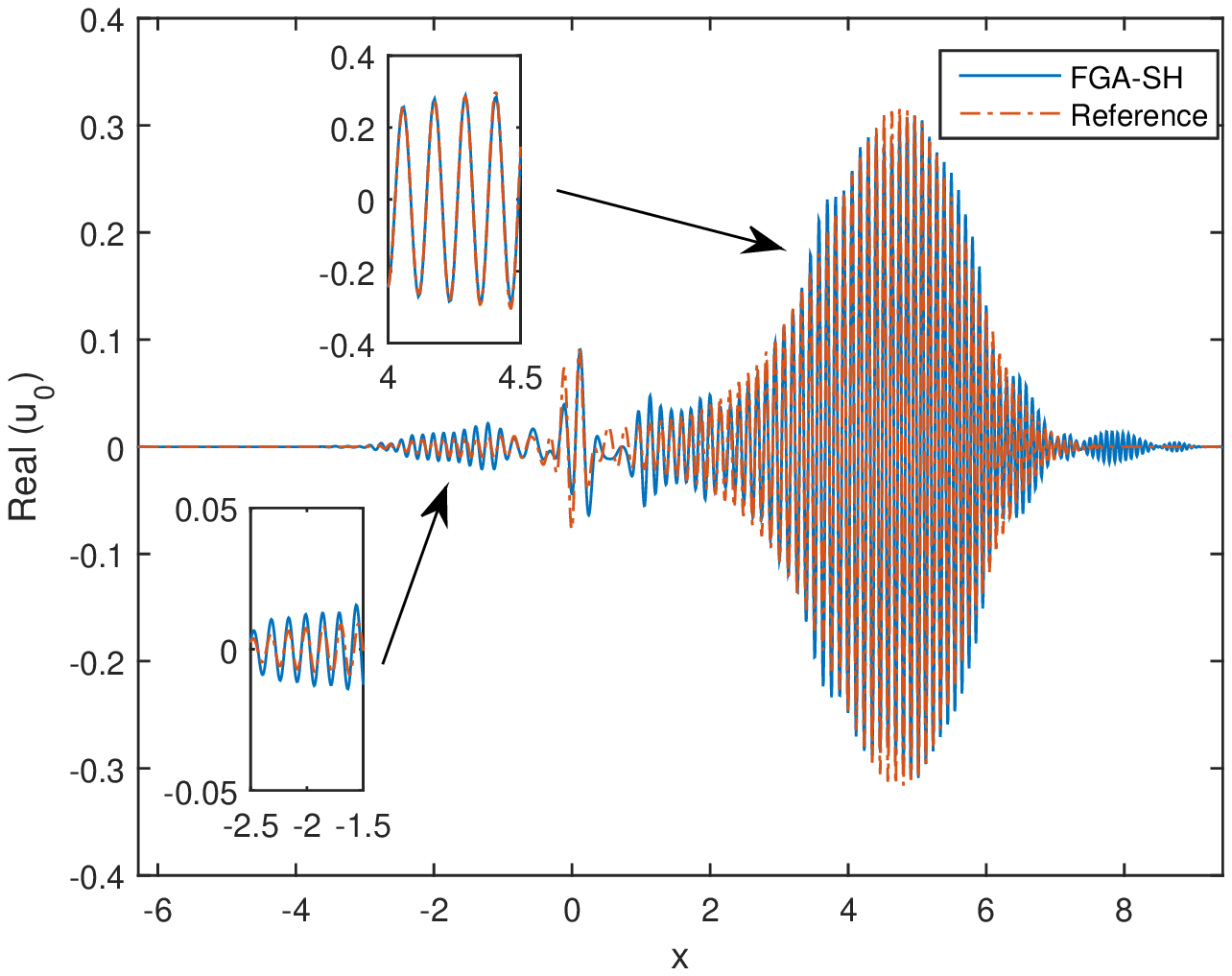}  \\
\includegraphics[scale=0.55]{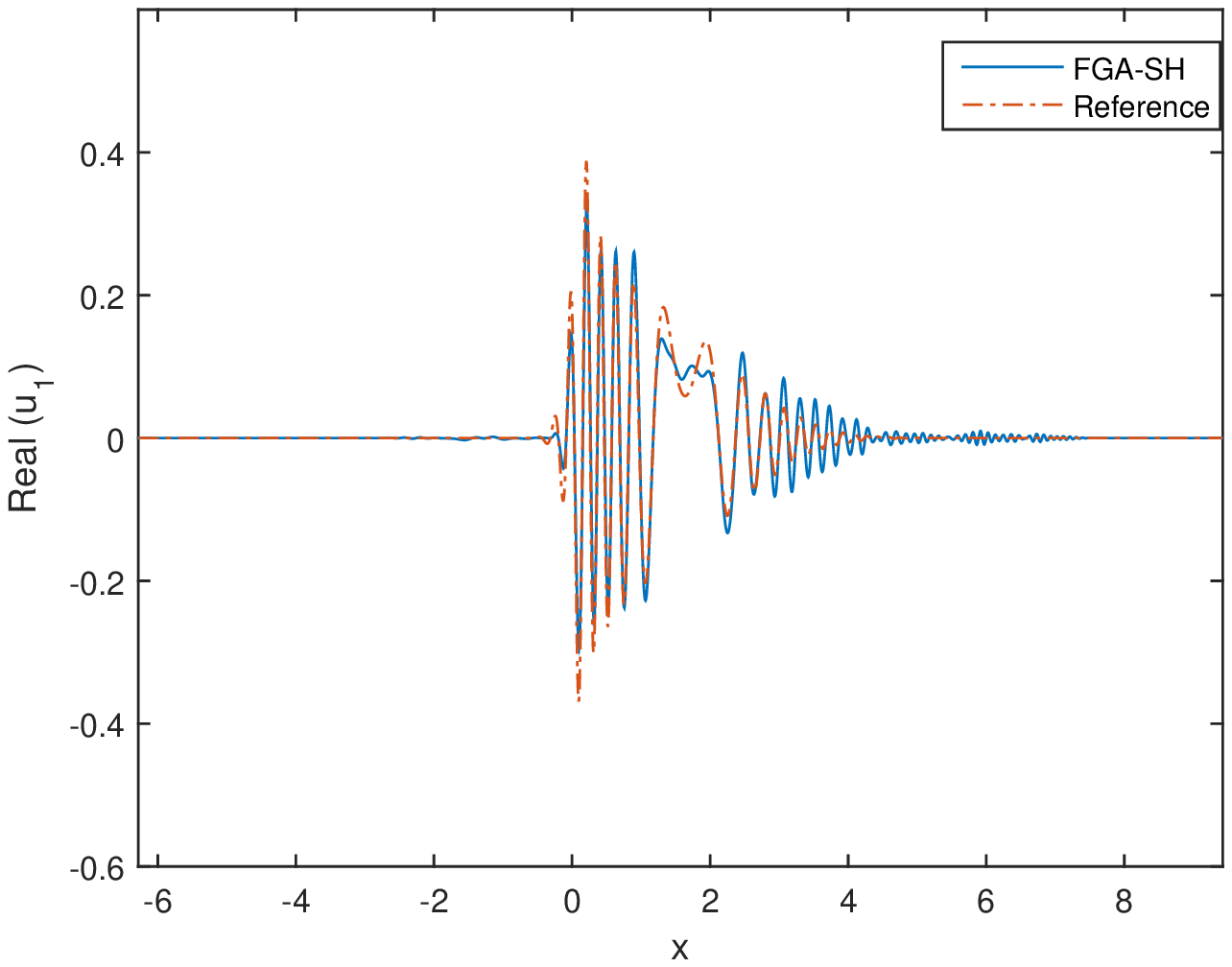}\\
\caption{Snapshot of the numerical solutions at $t=4$ (two components
  of the solution are plotted), note in particular that part of the
  wave function has bounced back and propagates to the left.}
\label{fig:test2-1}
\end{figure}

We now study the performance of the FGA-SH algorithm in a bouncing
back test, to validate the long time accuracy especially when the wave
propagation switches direction (bounces back by the energy
surface). We choose to focus on Example 1 for the test. We fix
$\veps=\frac{1}{32}$, $w=2$, $\delta=\veps$ and $C_g=5$. The potential
is steeper compared to that of the previous subsection, and therefore
the wave packet tends to be bounced back by the steep potential. The
initial condition is chosen concentrated on the lower surface only,
and takes the same form as \eqref{eq:u00} with $k_0=1.7$ and
$y_0=-1.5$. We choose these parameters so that the transmitted
  wave packet will switch direction when propagating on the top energy
  surface and experience the second significant non-adiabatic
transition when it goes through $x = 0$ again.  We compare the
numerical results with initial number of trajectory $M= 3200$ to the
reference solution till $T=4$. The wave functions obtained by the
FGA-SH method with reference solutions are plotted in Figure
\ref{fig:test2-1}, from which we observe very nice agreement.

\begin{figure}
\includegraphics[scale=0.575]{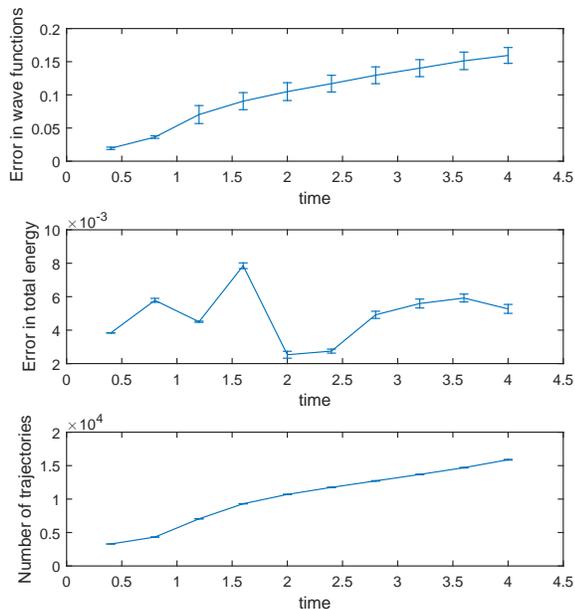}  \\
\caption{From top to bottom: The error in the wave functions measured
  in $L^2$ metric, the error in total energy, and the number of
  trajectories as a function of time. The averages and confidence
  intervals are estimated empirically using $100$ runs.}
\label{fig:test2-2}
\end{figure}

We plot the error in the wave function versus time in Figure
\ref{fig:test2-2} (the error is empirically averaged for independent
run of $100$ times). It seems that the error grows roughly linearly in
time, so the growth is mild. Also, we plot the discrepancy of the
total energy between the FGA-SH solutions and the reference solutions.
In this test, the initial energy of the wave packet is given by
$0.1935$. Besides the averaged error shown in Figure
\ref{fig:test2-2}, we also note that among the $100$ runs, the largest
deviation in energy is $0.0115$. Therefore, the total energy is
roughly conserved even in the worst scenario (relative error
$5.94\%$). Also, we report the average number of trajectories versus
time in Figure \ref{fig:test2-2}, which suggests a mild linear growth
as the simulation proceeds. This is a good indication that the
variance of the algorithm does not grow rapidly in time, while
rigorous variance bound is beyond the scope of the current work, and
will be leaved for future works.

Let us remark that a major difference between FGA-SH and other FSSH
type algorithms is that the trajectory $\wt{z}(t)$ is continuous in
time on the phase space $\R^{2m}$, while in FSSH and other versions of
surface hopping, a momentum shift is usually introduced to conserve
the classical energy along the trajectory when hopping occurs (if
hopping occurs from energy surface $0$ to $1$, it is required that
$E_0(p, q) = E_1(p', q)$ where $p'$ is the momentum after hopping).
Note that as in the FGA for single surface Schr\"odinger equation,
each Gaussian evolved in the FGA-SH does not solve
the matrix Schr\"odinger equation, and only the average of
trajectories gives an approximation to the solution. Therefore, it is
not necessary for each trajectory to conserve the classical energy. As
we have shown, the ensemble average of the trajectories gives good
conservation of the energy of the wave packet.

\subsection{Effect of weighting factor $w$} \label{sec:weighting}

\begin{figure}
\includegraphics[scale=0.275]{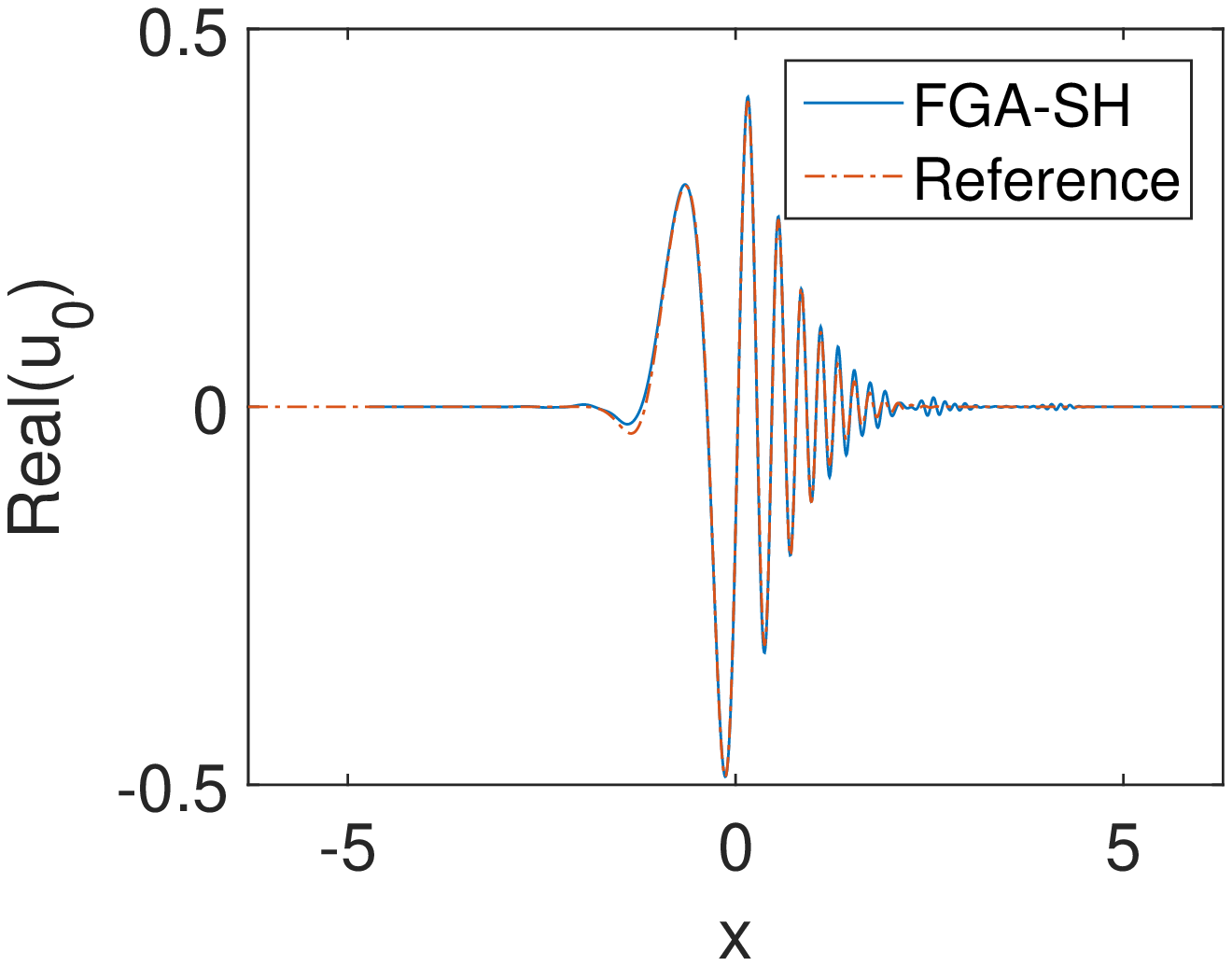} 
\includegraphics[scale=0.275]{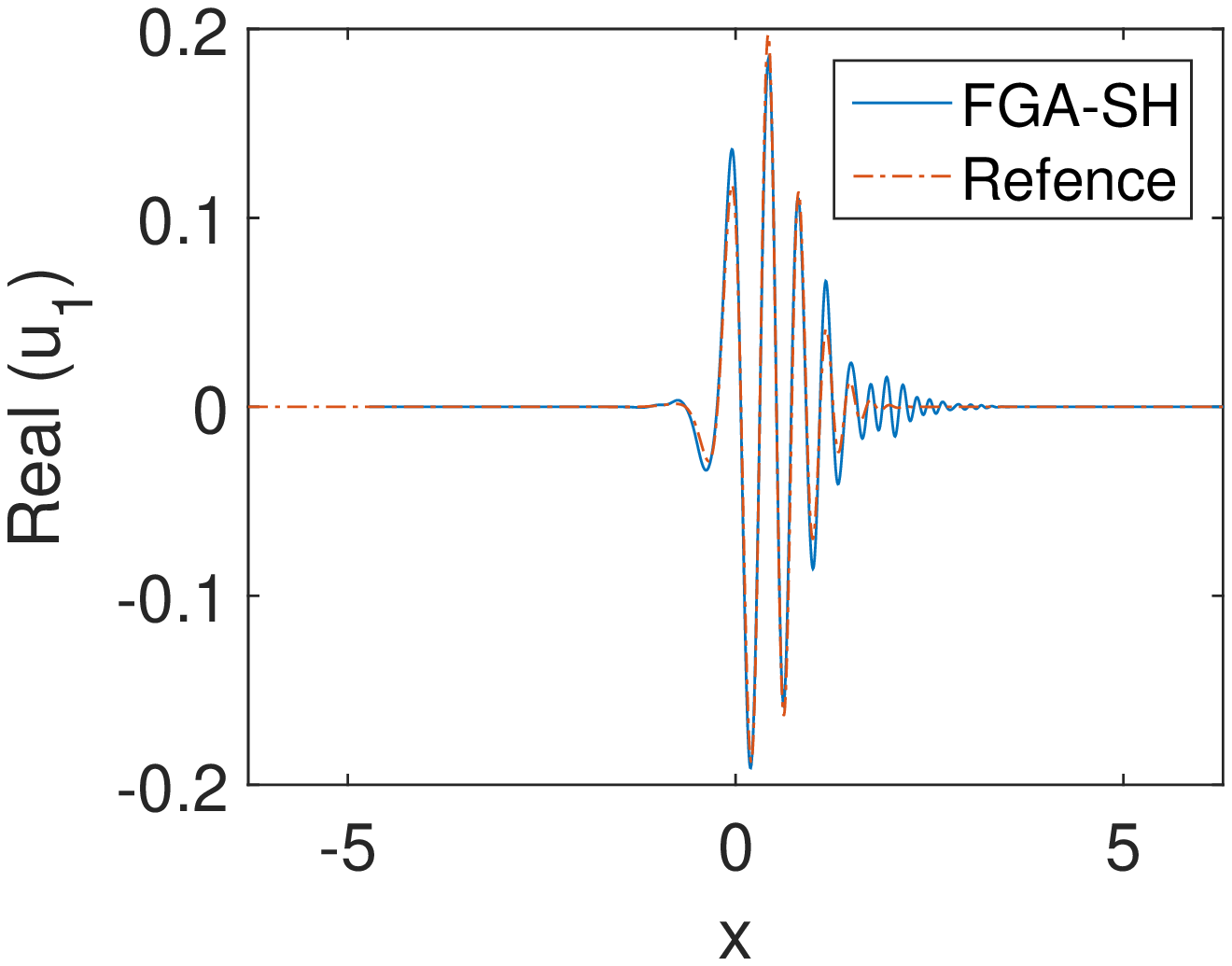} \\
\includegraphics[scale=0.275]{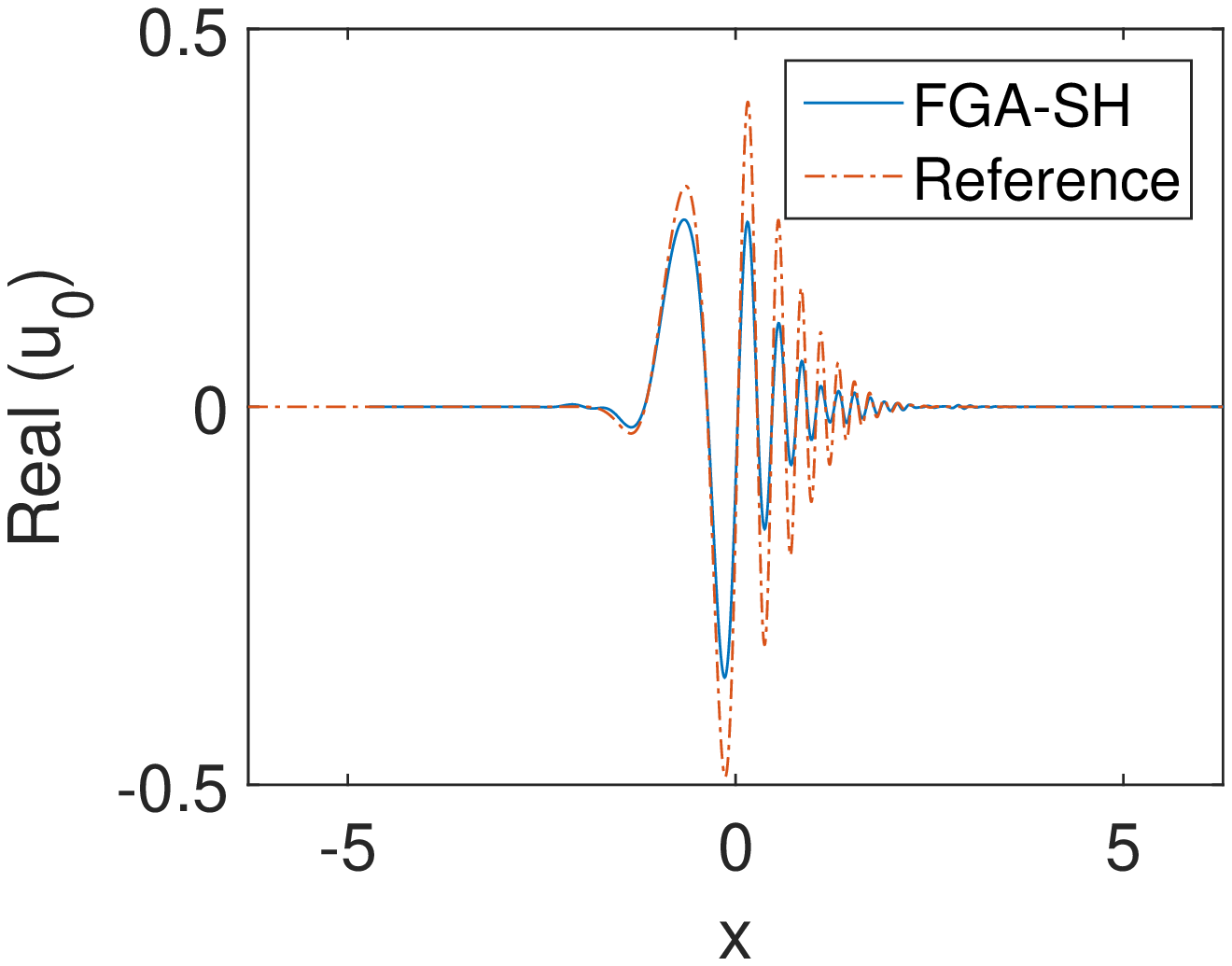}
\includegraphics[scale=0.275]{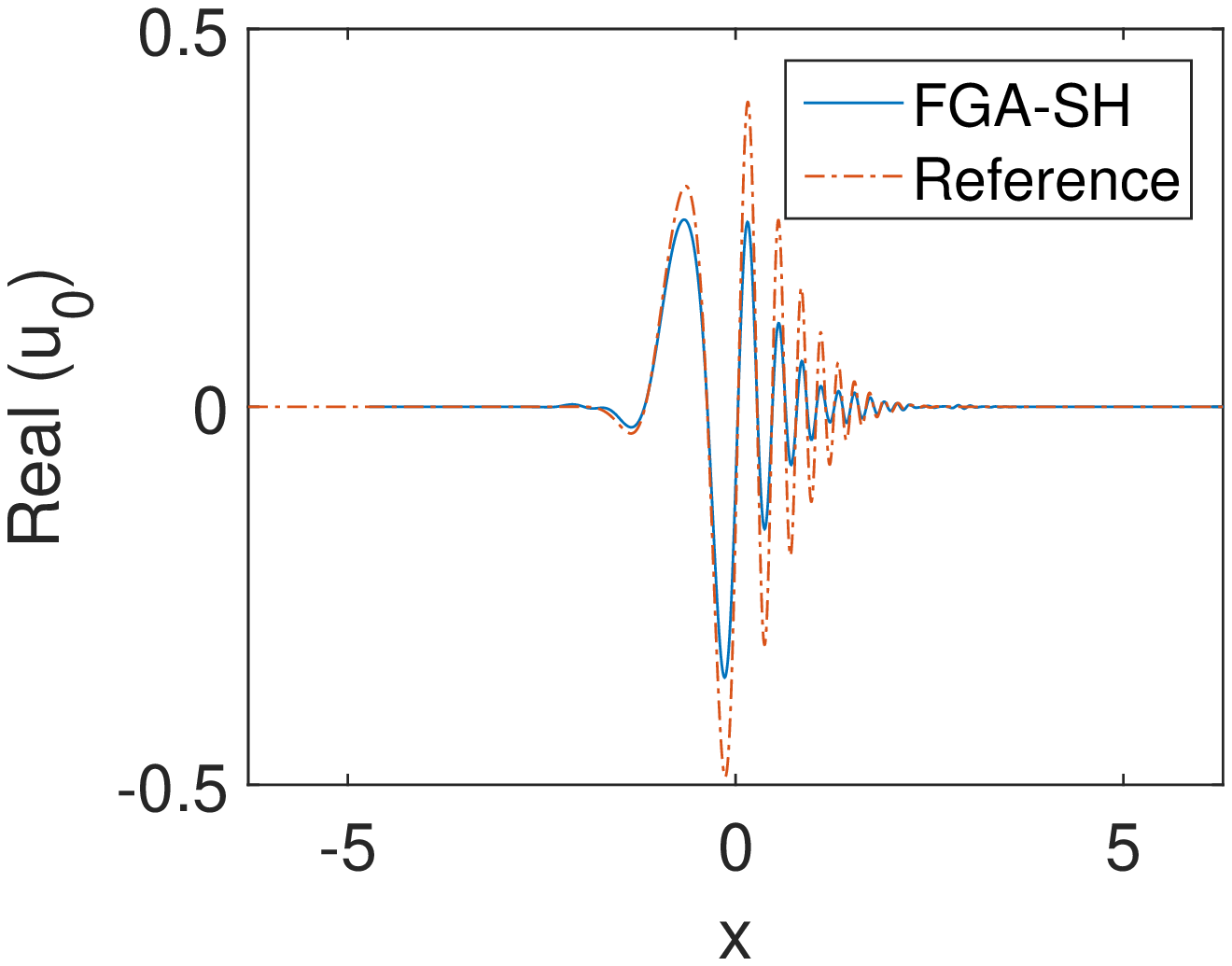}\\
\caption{Top: FGA-SH solutions with the weighting factor. Bottom: FGA-SH solutions without the weighting factor.}
\label{fig:test3}
\end{figure}

We now numerically demonstrate the role of the weighting factor
$\exp(w)$, which is important to get the correct ensemble average of
the wave function, since the hopping is driven by a non-homogenous
Poisson process (the jumping rate depends on the position and momentum of the trajectory). We choose to focus on Example 1 for
the test. We fix $\veps=\frac{1}{32}$, $w=1$, $\delta=\veps$ and
$C_g=1$.  The initial condition is chosen concentrated on the lower
surface only, and takes the same form as \eqref{eq:u00} with $k_0=1.5$
and $y_0=-1.5$. We run the algorithm with and without the weighting
factor $w$ (meaning setting always $w = 0$) for $100$ times each, some
snapshots of the numerical solutions are plotted in Figure
\ref{fig:test3}, from which we observe that, weighting factors are
crucial for accurate approximations of the wave functions.

Also, we summarize the empirical averages and the corresponding
variances of the wave functions error and the transition rates
(abbreviated by TR) in Table~\ref{table:test3}. We observe that the weighting factor $w$ is also
crucial to get good approximations of the transition rates.
\begin{table}
  \centering
  \begin{tabular}{ c| c| c|c|c} \hline
    & avg. error & Variance & TR mean & Variance  \\ \hline
    w/o w.f. & 0.2695 & 9.5780e-03 & 0.1772 & 1.0339e-02 
    \\ \hline
    with w.f. & 0.0678 & 7.5759e-03 & 0.2386 & 1.2106e-02 
    \\ \hline
\end{tabular}
\caption{Numerical error in the wave functions, and average transition rates with and without the weighting factor. The reference transition rate is given by $0.2443$; the inclusion of the weighting factor reduces the relative error from $27.5\%$ to $2.33\%$.}
  \label{table:test3}
\end{table}

%\subsection{Convergence for FSSH orAFSSH?}

\subsection{Initial condition with different momentum}

Finally, we test all three examples in Section \ref{sec:Tully} with
initial conditions of different momentum, we aim to compare the FGA-SH
method with reference solutions in terms of transition rates. We fixed
$\veps=\frac{1}{64}$ and initial sample size $M=1600$. The initial
condition is chosen concentrated on the lower surface only, and takes
the following wave packet form
\[
u_0(0,x)= (32\veps)^{-1/4} e^{\frac{i}{\veps}K \cdot (x-y_0)} e^{-
  \frac{1}{2\veps}(x-y_0)^2},
\]
where $y_0 = -1.5$ and various $K$ are used so that different momentum
is considered for the initial wave packet.

\begin{figure}[ht]
\includegraphics[scale=1]{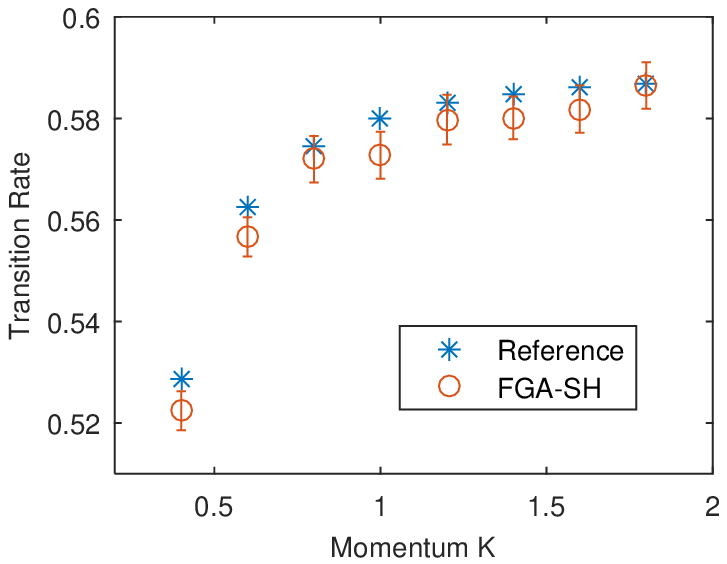} 
\includegraphics[scale=1]{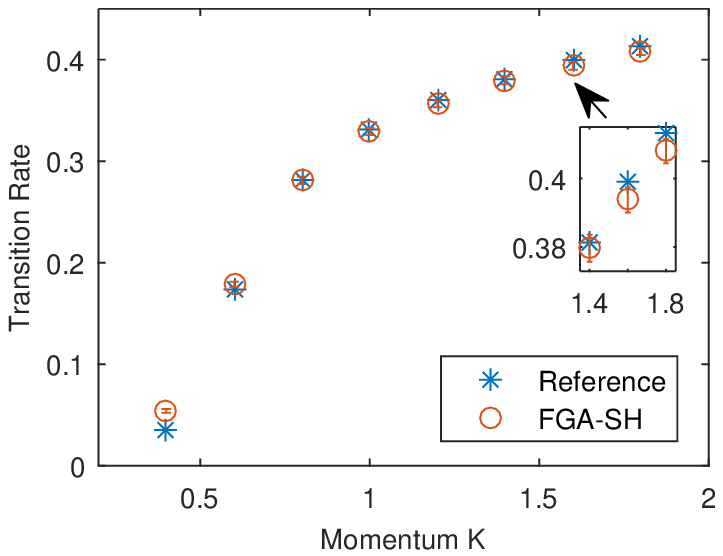} \\
\includegraphics[scale=0.275]{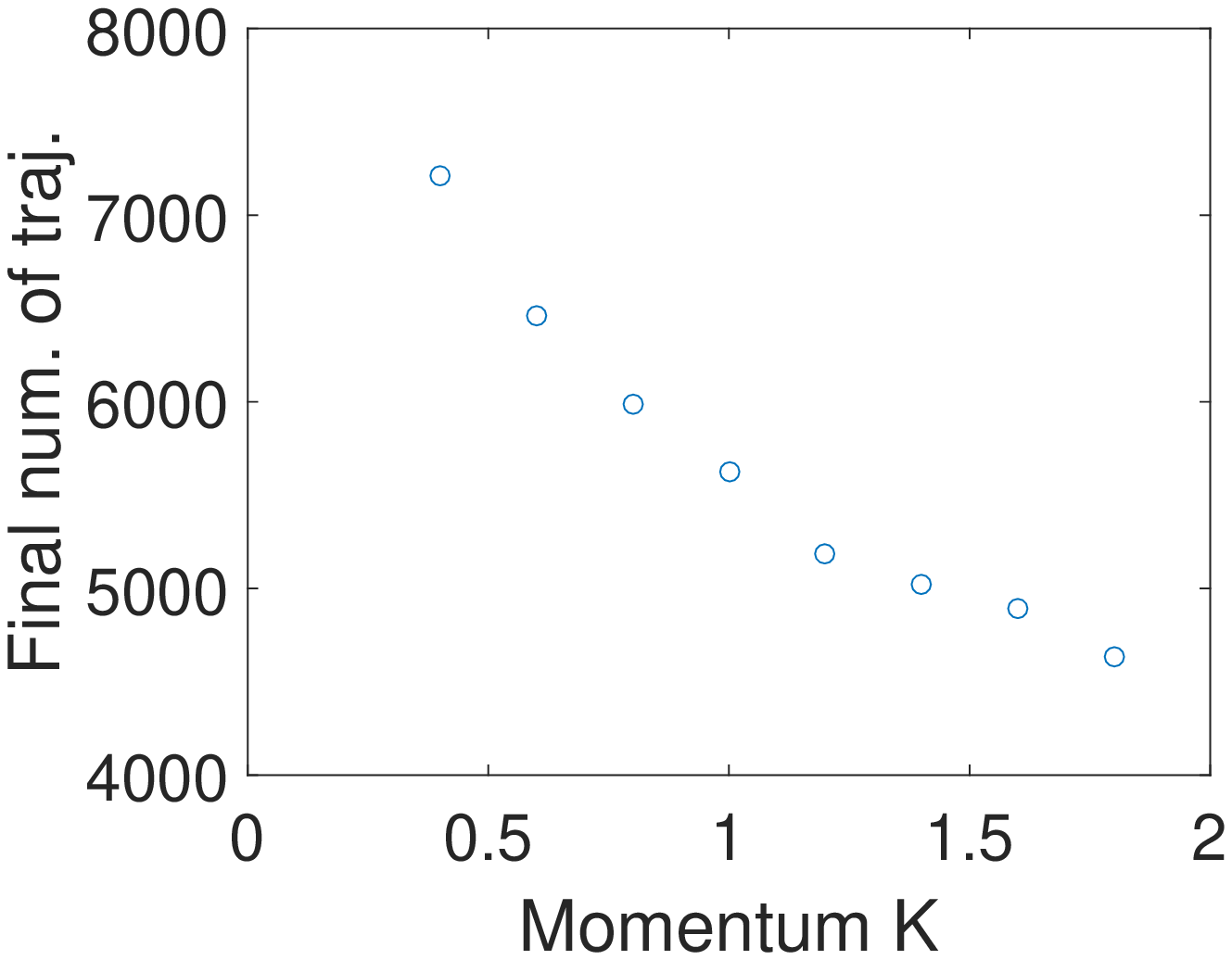} 
\includegraphics[scale=0.275]{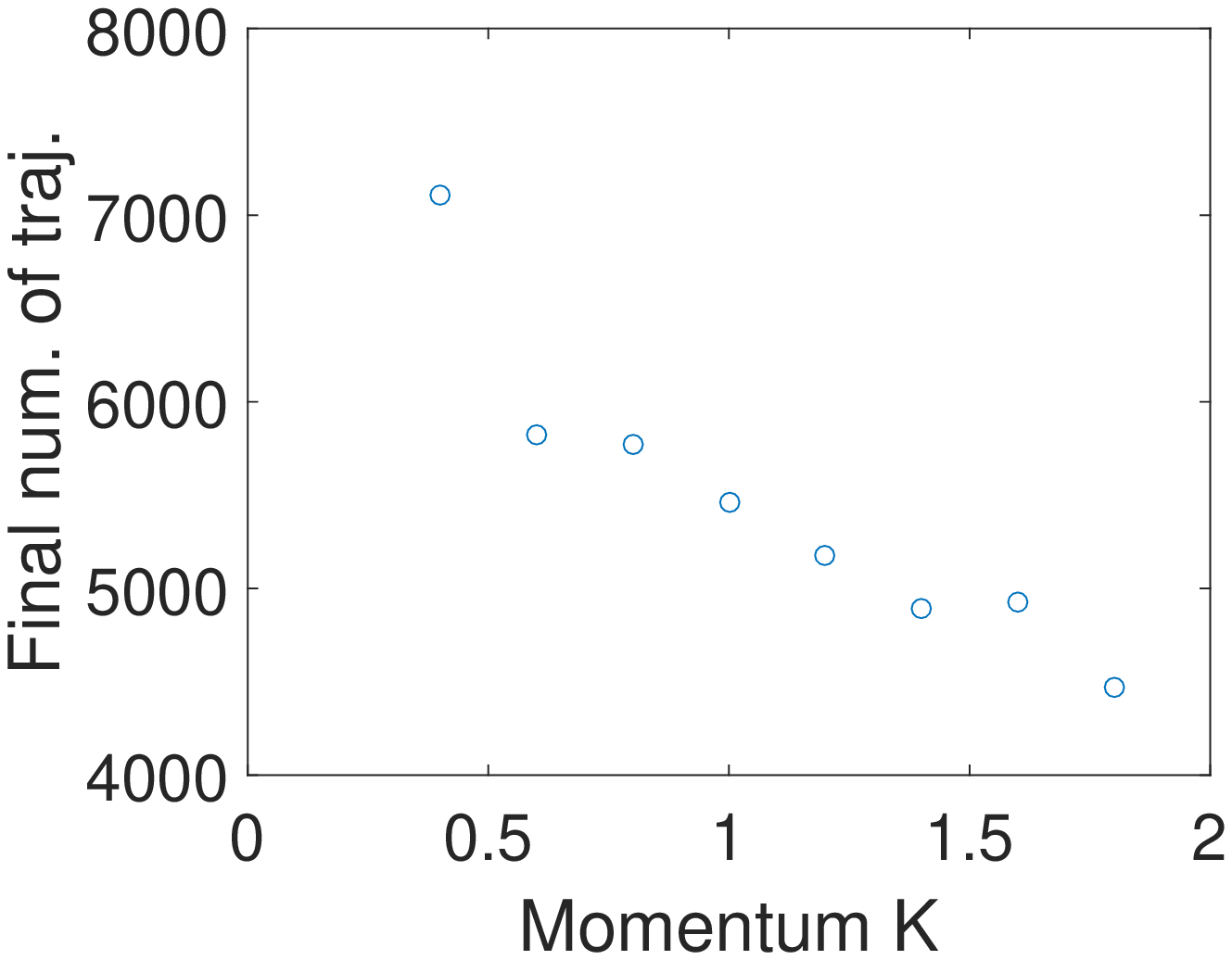} 
\caption{Numerical results of the transition rate of FGA-SH
    method compared with the reference solution for Example 1.
    Top: Transition rates for Example 1 with $C_g = \frac{1}{20}$
    (smaller gap); middle: Transition rates for Example 1 with
    $C_g = 1$ (larger gap); bottom left: typical number of
    trajectories at final time for Example 1 with small $C_g$;
    bottom right: typical number of trajectories at final time for
    Example 1 with large $C_g$.}
\label{fig:test4}
\end{figure}

\begin{figure}[ht]
\includegraphics[scale=1]{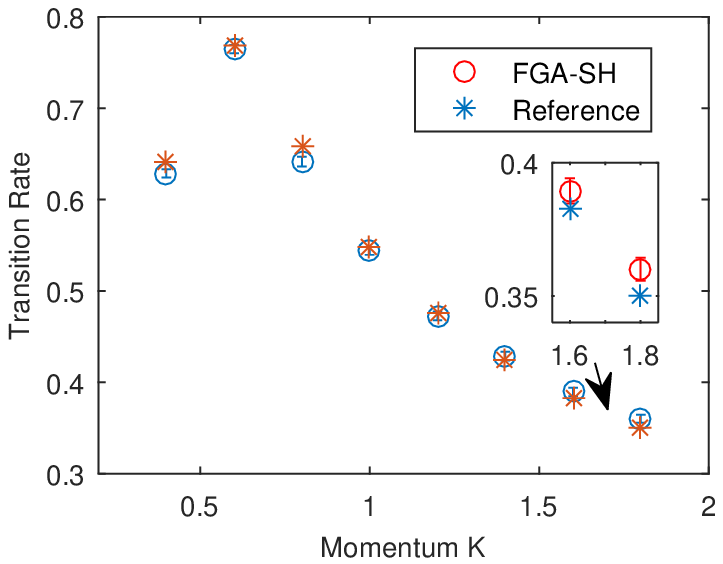}
\includegraphics[scale=1]{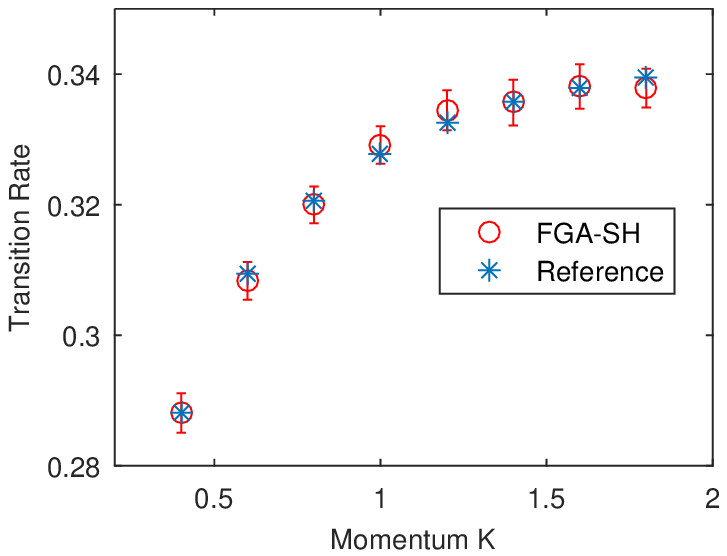}\\
\includegraphics[scale=0.275]{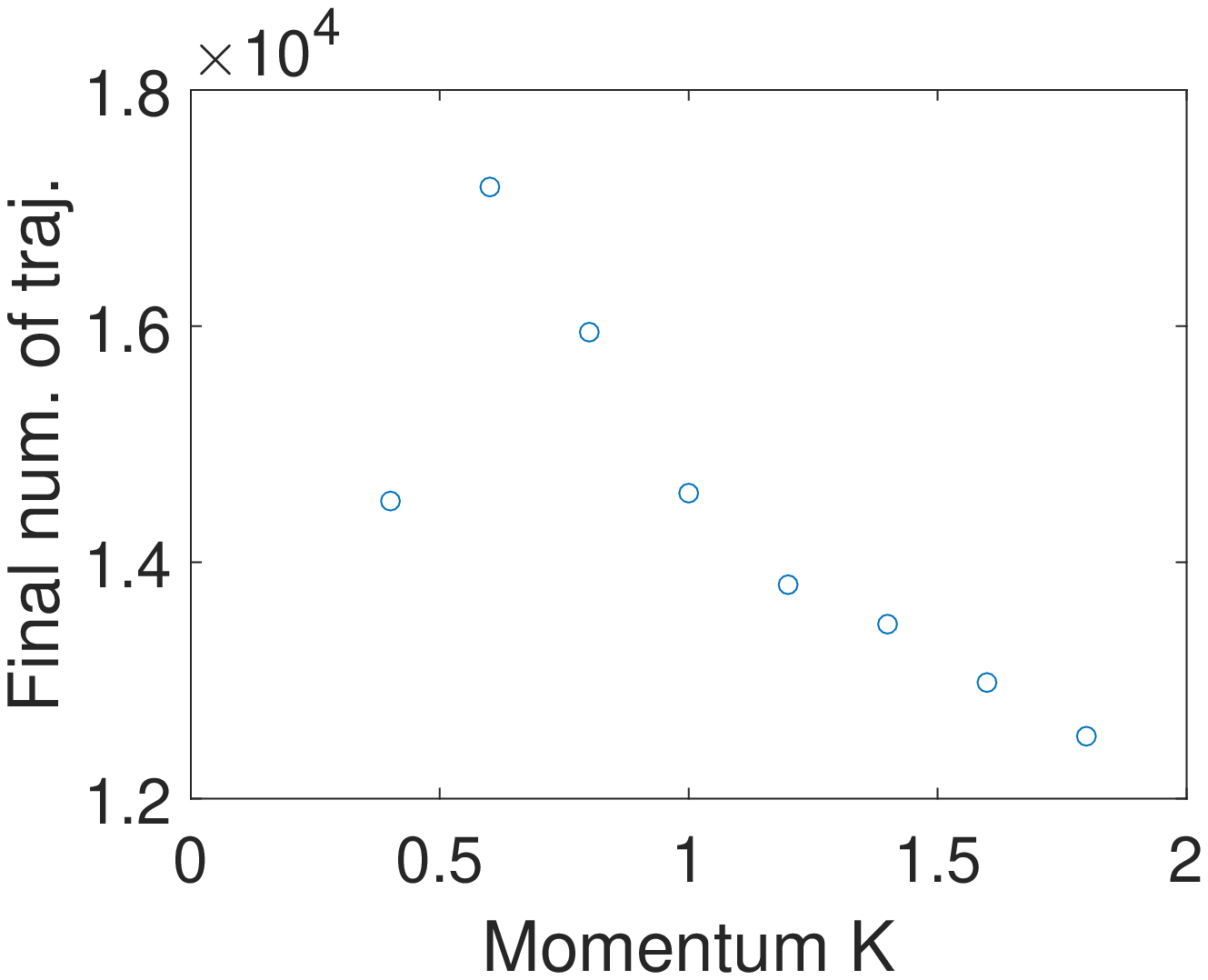} 
\includegraphics[scale=0.275]{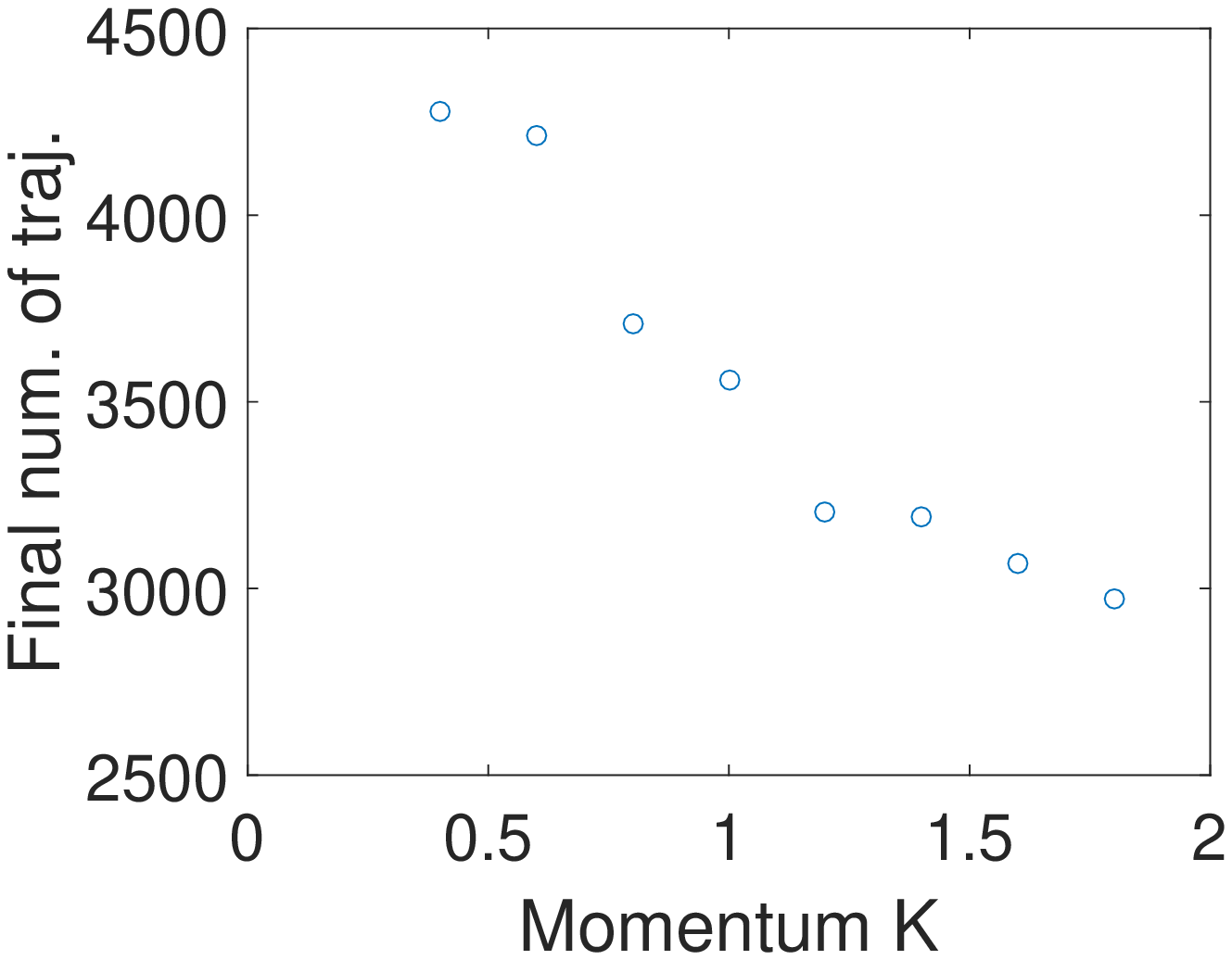} 
\caption{Numerical results of the transition rate of FGA-SH
    method compared with the reference solution for Example 2 and
    Example 3.  Top: Transition rates for Example 2; middle:
    Transition rates for Example 3; bottom left: typical number of
    trajectories at final time for Example 2; bottom right: typical
    number of trajectories at final time for Example 3.}
\label{fig:test5}
\end{figure}

For Example 1, we choose $\delta=5\veps$ and test two cases: the small
global gap scenario $C_g=\frac{1}{20}$ and the large gap scenario
$C_g=1$. Note that, when $C_g=1$, many classically forbidden hops will
happen.  The FGA-SH algorithm is repeated for $100$ independent trials
in each case, and empirical averages of the hopping rates with
confidence intervals are plotted in Figure~\ref{fig:test4}, together
with the typical number of trajectories at the end of the simulation
time. The corresponding results for Example 2 and Example 3 are
plotted in Figure~\ref{fig:test5}.  We observe that, the FGA-SH
results give accurate approximation in the tests. It is also worth
pointing out that the error seems to be rather uniform for different
values of the initial momentum $K$. We also remark that the
birth/death processes adaptively choose the number of trajectories
needed, which helps to maintain the uniform accuracy over different
initial momentum. From the numerical results, it can be seen that a
smaller initial momentum ends up requiring more trajectories.

%\begin{widetext}
%\end{widetext}

%%%%%%%%%%%
\section{Conclusion} \label{sec:conclusion}

In this work we further develop the FGA-SH method, introduced in
\cite{FGASH}, by proposing an improved sampling algorithm using birth
/ death branching processes.  The algorithm is validated in various
numerical tests for the standard test cases for non-adiabatic
dynamics. 

The path integral interpretation of the fewest switches surface
hopping type of algorithm leads to potentially further development for
algorithms for non-adiabatic dynamics. Some interesting future
directions include validation of the algorithm for higher dimensional
problems, non-adiabatic thermal sampling using surface hopping
dynamics and also the calculation time-correlation function in the
non-adiabatic regime.

\appendix

%%%%%%%%%
\section{Asymptotic derivation of the path integral semiclassical
  approximation}

For completeness, we provide a brief explanation of the path integral
approximation \eqref{eq:FGApathintegral}. Please refer to \cite{FGASH}
for the detailed asymptotic derivation and mathematical proofs.  For
simplicity of notation, we assume the initial condition is on the
energy surface $E_0$, and hence the trajectory starts from that energy
surface.

We assume the following deterministic ansatz, referred as surface
hopping ansatz, for the solution to \eqref{vSE}.
\begin{multline}\label{eq:ufga}
  u_{\FGA}(T,x)=\ket{0} \left(u^{(0)}(T, x)+u^{(2)}(T, x)+\cdots\right) \\
  + \ket{1} \left(u^{(1)}(T, x)+u^{(3)}(T, x)+\cdots\right).
\end{multline}
This ansatz is similar to that of proposed by Wu and Herman
\cite{WuHerman:07,WuHerman:06, WuHerman:05}, which is also based on
the Herman-Kluk propagators. The two approaches are different however
in several essential ways, as elaborated in \cite{FGASH}.

The wave function $u^{(n)}$ stands for the contribution with $n$
surface hops before time $t$, starting from surface $E_0$. In
  particular, for trajectories with even number of hops, the
  electronic state ends at $\ket{0}$, and trajectories with odd number
  of hops contribute to $\ket{1}$. This explains the linear
  combination in \eqref{eq:ufga}.  We denote a sequence
$\{t_k\}_{k=1}^n$ for the hopping times satisfying
\[
0 \le t_1 \le t_2 \le \cdots \le t_n \le T,
\]
at which time the trajectory switches from one energy surface to the
other. The ansatz for $u^{(n)}$ is given by
\begin{multline}\label{eq:u0n}
  u^{(n)}(T, x) =  \frac{1}{(2\pi\veps)^{3m/2}} \int \ud z_0  \int_{0 \le t_1 \le\cdots \le t_n \le T} \\
  \tau^{(1)}\cdots\tau^{(n)} \; A^{(n)} \; \exp\left( \frac{i}{\veps}
    \Theta^{(n)}\right) \ud T_{n:1},
\end{multline}
where $\tau^{(k)}$ is defined in \eqref{eq:deftau} and
$\ud T_{n:1}=\ud t_1 \cdots \ud t_n$. Note that in \eqref{eq:u0n}, we
integrate over all possible hopping times for $n$ hops in the time
interval $[0, T]$. Given $\{t_k\}_{k=1}^n$ and $z_0$, the trajectory
$\wt {z}(t)$ for $0 \le t\le T$ is specified.

Substitute the ansatz into the matrix Schr\"odinger equations and
carry out asymptotic calculations as in \cite{FGASH}, we arrive at the
conclusion that the evolution of $A^{(n)} $ and $ \Theta^{(n)}$ should
be exactly as that described in Section \ref{sec:functional}, such
that $u_{\FGA}(T, x)$ is a good approximation to the true
solution. Indeed, the asymptotic analysis can be turned into rigorous
error analysis \cite{FGASH} that, $u_{\FGA}(T,x)$ is an approximation
of the exact solution with $\mathcal O (\veps)$ error in $L^2$ metric.

Let us now link the deterministic ansatz to the path integral
representation.  As we discussed in Section \ref{sec:trajectory},
given $T>0$, the number of jumps $n$ of the stochastic trajectory
$\wt {z}(t)$ for $0 \le t\le T$ is a random variable.  In particular,
by the properties of the associated counting process, the probability
that there is no jump $(n = 0)$ is given by
\begin{equation}
  \mathbb{P}(n = 0) 
  = e^{-\int_0^T \Abs{\tau^{(1)}} \ud s }. 
\end{equation}
And, more generally, we have
\begin{multline}\label{eq:Pnk}
  \mathbb{P}(n = k) = \int_{0<t_1<\cdots<t_k<T} \ud T_{k:1} \; \prod_{j=1}^k \Abs{\tau^{(j)}} \\
  \times e^{-\int_{t_k}^t \Abs{\tau^{(k+1)}} \ud s } \prod_{j=1}^{k}
  e^{ - \int_{t_{j-1}}^{t_j} \Abs{\tau^{(j)}} \ud s },
\end{multline}
and the probability density of $(t_1, \cdots, t_k)$ given there are
$k$ jumps in total is
\begin{multline}\label{eq:varrhok}
  \varrho_k(t_1, \cdots, t_k) \propto
  e^{-\int_{t_k}^t \Abs{\tau^{(k+1)}} \ud s  } \prod_{j=1}^k \Abs{\tau^{(j)}} \times \\
  \times \prod_{j=1}^{k} e^{ - \int_{t_{j-1}}^{t_j} \Abs{\tau^{(j)}}
    \ud s },
\end{multline}
for $t_1 \le t_2 \le \cdots \le t_n$, and $0$ otherwise.

Using the above probabilities, we may calculate explicitly the
expectation with respect to the trajectory $\wt{z}$. We verify that
\eqref{eq:FGApathintegral} is exactly a stochastic representation of
the FGA ansatz given in \eqref{eq:ufga}--\eqref{eq:u0n}, where the
integrals with respect to $t_1, \ldots, t_n$ are replaced by the
averaging of trajectories. In particular, for the functional
  $\mc{F}$ \eqref{eq:defF}, we observe that the term
  $A(T) \exp\Bigl(\frac{i}{\veps} \Theta(T, x)\Bigr)$ comes from the
  integrand in \eqref{eq:u0n}. The weight factor
  $\exp\bigl(w(T)\bigr)$ comes from the probability of $k$ hops as in
  \eqref{eq:varrhok}. The appearance of ratios like
  $\tau^{(k)}/ \abs{\tau^{(k)}}$ in \eqref{eq:defF} is to match
  $\tau^{(k)}$ in the integrand \eqref{eq:u0n} with the use of
  $\abs{\tau^{(k)}}$ in the hopping probability and hence in the
  expression \eqref{eq:varrhok}.  

%%%%%%%%%%%%%%%%%%%%
\begin{acknowledgments}
  This work is partially supported by the National Science Foundation
  under grants DMS-1454939 and RNMS11-07444 (KI-Net). J.L. would like
  to thank Joseph Subotnik and John Tully for helpful discussions.
\end{acknowledgments}

\bibliography{surfacehopping} 
\end{document}